\documentclass[a4paper,11pt]{article}
\usepackage{jheppub}
\usepackage{lineno}
\usepackage{multirow}
\usepackage{subcaption}
\usepackage{tabularx}
\usepackage{tikz}
\usepackage{makecell}  %% For multi-line cells in tables

\title{\boldmath Detecting highly collimated photon-jets from Higgs boson exotic decays with deep learning}

\author[1]{Xiaocong Ai} 
\author[2]{William Y. Feng}
\author[2]{Shih-Chieh Hsu}
\author[2]{Ke Li}
\author[3]{Chih-Ting Lu}
\affiliation[1]{School of Physics and Microelectronics, Zhengzhou University, Zhengzhou, 450001, China}
\affiliation[2]{Department of Physics, University of Washington at Seattle, \\ Seattle, Washington 98195, USA}
\affiliation[3]{Department of Physics and Institute of Theoretical Physics, \\ Nanjing Normal University, Nanjing, 210023, China}

\abstract{
    Recently, there has been a growing focus on the search for anomalous objects beyond standard model (BSM) signatures at the Large Hadron Collider (LHC). This study investigates novel signatures involving highly collimated photons, referred to as photon-jets. These photon-jets can be generated from highly boosted BSM particles that decay into two or more collimated photons in the final state. Since these photons cannot be isolated from each other, they are treated as a single jet-like object rather than a multi-photon signature. The Higgs portal model is utilized as a prototype for studying photon-jet signatures. Specifically, GEANT4 is employed to simulate electromagnetic showers in an ATLAS-like electromagnetic calorimeter, and three machine learning techniques: Boosted Decision Trees (BDT), Convolutional Neural Networks (CNN), and Particle Flow Networks (PFN) are applied to effectively distinguish these photon-jet signatures from single photons and neutral pions within the SM backgrounds. Our models attain an identification efficiency exceeding $99\%$ for photon-jets, coupled with a rejection rate surpassing $99\%$ for SM backgrounds. Furthermore, the sensitivities for searching photon-jet signatures from the Higgs boson exotic decays at the High-Luminosity LHC are obtained.
}

\emailAdd{xiaocongai@zzu.edu.cn, wyf1729@uw.edu, schsu@uw.edu, ke.li@cern.ch, ctlu@njnu.edu.cn}

\begin{document}
\maketitle
\flushbottom

\section{Introduction}

%\begin{itemize}
%    \item Motivation about why we need to apply ML and the improved results after using ML (review of similar studies).  Todo: Shih-Chieh
%\end{itemize}

The Higgs boson has held a position of central importance for over a decade since its discovery~\cite{CMS:2022dwd,ATLAS:2022vkf}. While an increasing body of precise measurements has lent substantial support to its compatibility with the Standard Model (SM) predictions, a fundamental question still looms large: Is this particle indeed the SM Higgs boson, or might it be a SM-like counterpart, affiliated with models beyond the SM? Such inquiries have led to the exploration of Higgs portal models, representing a bridge between the SM sector and the dark sector~\cite{Arcadi:2019lka}. Higgs portal models can address diverse and pressing issues including neutrino mass and mixing~\cite{DeRomeri:2020wng}, the nature of dark matter (DM)~\cite{Arcadi:2019lka,Matsumoto:2018acr,Bondarenko:2019vrb}, the electroweak phase transition~\cite{Cohen:2012zza,Huang:2012wn,Chao:2014ina}, cosmic inflation~\cite{Bezrukov:2009yw,Bezrukov:2021mio}, and the hierarchy problem~\cite{Batell:2022pzc}. In the quest to unravel the mysteries of these Higgs portal models, significant efforts are made not only in precision measurements of Higgs boson couplings to SM particles, but also in exploring its invisible and exotic decay channels~\cite{Shrock:1982kd,Curtin:2013fra}, which provide crucial insights into previously uncharted territories.

As investigations at the Large Hadron Collider (LHC) have continued, the absence of concrete evidence for new physics beyond the SM has left the scientific community pondering a crucial possibility: perhaps conventional search strategies, which have guided our exploration of new physics at the LHC thus far, represent merely a subset of the manifold ways to unveil the enigmas of new physics. New physics signal signatures may prove to be more intricate and unconventional than previously imagined. In response to this evolving landscape, the pursuit of anomalous objects beyond the SM has garnered growing interest within the scientific community~\cite{Chakraborty:2017mbz,Alimena:2019zri,Albouy:2022cin,Franceschini:2022vck}. Many of these searches for anomalous objects defy the traditional methodologies used in the past, necessitating the development of innovative search strategies and the deployment of cutting-edge techniques. This shift in focus represents a profound and exciting frontier in the ongoing journey of unraveling the mysteries of new physics.

Within the exploration of these anomalous objects, our focus centers on a unique class of signal signatures: photon-jets~\cite{Dobrescu:2000jt,Toro:2012sv}. Photon-jets emerge as the byproducts of the decay of new, highly boosted light particles, resulting in the generation of two or more closely collimated photons in their final state. Their peculiarity lies in the fact that these constituent photons cannot pass the conventional photon isolation criteria, leading us to treat them as a jet-like object, instead of the multi-photon signature~\cite{Steinberg:2021iay,Alves:2021puo,Knapen:2021elo}. The jet substructure analysis has been applied to this novel photon-jet signature~\cite{Ellis:2012sd,Ellis:2012zp,Chakraborty:2017mbz,Wang:2021uyb}. On the other hand, both ATLAS and CMS Collaborations have sought photon-jet signatures using the traditional cut-and-count method and Boosted Decision Trees (BDT) techniques~\cite{ATLAS:2012soa,Aaboud:2018djx,CMS:2022wpu,ATLAS:2023eet}. However, as the structure within the photon-jet becomes more intricate, deep learning techniques have proven more powerful than previous analysis strategies.

While previous studies have explored the application of Convolutional Neural Networks (CNN) in the context of a simple scenario involving only two closely collimated photons within the photon-jet signature~\cite{Ren:2021prq,Wang:2023pqx}, our ambition extends further. We aim to delve into the realm of more intricate photon-jet structures, specifically exploring configurations with two, four, and six closely collimated photons. This endeavor is undertaken with the Higgs portal model as our concrete prototype, allowing us to explore photon-jet signatures within this framework. In particular, we consider the exotic decays of the Higgs boson into a pair of light scalars (dark Higgs bosons)~\cite{Toro:2012sv,Ellis:2012sd,Ellis:2012zp,Chang:2015sdy,Sheff:2020jyw} or light pseudoscalars (axion-like particles)~\cite{Dobrescu:2000jt,Toro:2012sv,Draper:2012xt,Aparicio:2016iwr,Ellwanger:2016qax,Domingo:2016unq,Chiang:2016eav,Domingo:2016yih,Wang:2021uyb,Ren:2021prq,Lu:2022zbe,Lane:2023eno}. There is a substantial mass difference between the scalar/pseudoscalar and the Higgs boson, leading to the production of these light particles with high momentum and, consequently, the formation of photon-jets when they predominantly decay into multiple photons in the final state. To effectively distinguish these photon-jet signatures from single photons and neutral pions within the SM backgrounds, we employ three machine learning techniques: BDT~\cite{Roe:2004na}, CNN~\cite{Ayyar:2020ijy}, and Particle Flow Network (PFN)~\cite{Komiske:2018cqr}. This approach extends beyond conventional boundaries and offers a complementary method to new light particle searches conducted at B-factory and fixed target experiments~\cite{Filimonova:2019tuy,Belle-II:2020jti,Ferber:2022rsf,Belle-II:2023ueh,Blumlein:1990ay,Dobrich:2015jyk,Feng:2018pew,Harland-Lang:2019zur,Gorbunov:2021ccu,Kling:2022uzy,Liu:2023bby}.

The structure of this paper is organized as follows. In Sec.~\ref{sec:model}, we provide a concise review of the Higgs portal model, focusing on the photon-jet signatures within this framework. The application of three distinct machine learning techniques to the identification of photon-jets and a demonstration of their respective performances are presented in Sec.~\ref{sec:identification}. We discuss the obtained results and their implications for the Higgs portal model in Sec.~\ref{sec:results}. Finally, a summary of our findings is presented in Sec.~\ref{sec:conclusion}.
\section{The Higgs portal model} 
\label{sec:model}

In this study, we investigate a simplified Higgs portal model that extends beyond the SM. This model introduces a framework in which, alongside the SM-like Higgs boson denoted as $h_1$ with $m_{h_1}\simeq 125$ GeV, we consider the presence of either a light scalar, $h_2$, or a light pseudoscalar, $a$. These additional particles, namely $h_2$ and $a$, play several roles within the model. The light scalar can potentially serve as a mediator for sub-GeV DM models~\cite{Matsumoto:2018acr,Bondarenko:2019vrb}, a candidate for the light inflaton~\cite{Bezrukov:2009yw,Bezrukov:2021mio}. Moreover, a singlet complex scalar field is often used to provide the mass of dark photon for dark sector models with extra $U(1)$ gauge symmetry, employing the dark Higgs mechanism~\cite{Baek:2014kna}. Similarly, the light pseudoscalar can manifest as an axion-like particle, serving as a mediator for sub-GeV DM models~\cite{Dolan:2014ska,Dolan:2017osp,Bharucha:2022lty,Ghosh:2023tyz}, and even offering a solution to experimental anomalies, such as the muon $g-2$ excess~\cite{Liu:2022tqn} and the MiniBooNE excess~\cite{Chang:2021myh}, among others.

In this framework, the spin-$0$ particles ($h_1$, $h_2$, and $a$) can interact with one another or themselves via three-point and four-point interactions. Our particular focus in this investigation centers on exploring the $h_1$-$h_2$-$h_2$ and $h_1$-$a$-$a$ interactions, characterized by the couplings $\mu_{h_1 h_2 h_2}$ and $\mu_{h_1 aa}$, which have the same dimension as mass. Based on the motivations outlined above, encompassing light scalar and pseduoscalar, and with the aim of studying novel photon-jet signatures, we concentrate on the sub-GeV scale $h_2$ and $a$ in this study~\cite{Bauer:2018uxu,Winkler:2018qyg}. Moreover, we assume that $h_2\rightarrow aa$ is kinematically forbidden for the sake of simplicity. Concretely, the partial decay width for $h_1 \to h_2 h_2$ can be represented as  
\begin{equation}
\Gamma (h_1\to h_2 h_2) = \frac{\mu^2_{h_1 h_2 h_2}}{32\pi m_{h_1}}\sqrt{1-4\left(\frac{m_{h_2}}{m_{h_1}}\right)^2}.  
\label{Eq:h1_exotic}
\end{equation} 
The partial decay width for $h_1 \to aa$ is the same as Eq.~(\ref{Eq:h1_exotic}) except for changing the notation $h_2$ to $a$.

In our investigation, we concentrate on final states characterized by multiple photons and explore the charge-parity properties of light scalars and pseudoscalars, with a specific focus on pairs of $h_2$ or $a$ produced from $h_1$. These secondary particles then undergo distinct decay processes: $h_2$ decays into $\gamma\gamma$ and $\pi^0\pi^0$~\cite{Ellis:2012sd,Ellis:2012zp,Chang:2015sdy}, while $a$ transforms into $\gamma\gamma$, $\pi^0\pi^0\pi^0$~\cite{Dobrescu:2000jt}, where the $\pi^0$ subsequently decays into a pair of photons due to the chiral anomaly effect. To highlight these photon-rich decay modes from $h_2$ and $a$, we consider a mass range of approximately $0.45$ GeV $\lesssim m_{h_2, a}\lesssim 1.0$ GeV, ensuring their dominance in the overall decay processes.

In our study, we illustrate the distinctive characteristics of the Higgs boson exotic decay process, specifically $h_1\rightarrow h_2 h_2/aa\rightarrow 4\gamma$. An essential aspect of our investigation involves the determination of the opening angle between two collimated photons arising from the decay of either $h_2$ or $a$. This opening angle, denoted as $\Delta R_{\gamma\gamma}$, can be estimated as $\Delta R_{\gamma\gamma}\sim \frac{2}{\gamma_{h_2, a}}\sim 4\frac{m_{h_2, a}}{m_{h_1}}$, and it falls within the range of $0.015$ to $0.035$ for $0.45$ GeV $\lesssim m_{h_2, a}\lesssim 1.0$ GeV. Here, $\gamma_{h_2, a}\sim\frac{m_{h_1}}{2m_{h_2, a}}$ denotes the Lorentz factor of $h_2$ or $a$. Significantly, $\Delta R_{\gamma\gamma}\lesssim 0.04$ is approximately the same size as a typical energy cluster resulting from a single photon within the ATLAS electromagnetic calorimeter (ECAL)~\cite{ATLAS:2018dfo}. Consequently, the small opening angle between two photons in our analysis poses a challenge for the existing ATLAS triggers, as they cannot readily distinguish between an energy deposit in the ECAL arising from the photon-jet and that from a single photon. Therefore, at the ATLAS trigger level, our photon-jet is initially identified as a single photon-like object and subsequently recorded for further analysis. The primary objective of our work is to effectively differentiate such photon-jets from single photons or neutral pions within the ATLAS ECAL using advanced machine learning techniques. However, when $\Delta R_{\gamma\gamma}\gtrsim 0.3$, the two photons can be treated as two isolated photons within the ATLAS detector~\cite{ATLAS:2015rsn}. In general, photons with opening angles within the range of $0.04\lesssim\Delta R_{\gamma\gamma}\lesssim 0.3$ can be categorized as photon-jet objects. Unlike previous studies on photon-jets that consider relatively larger $\Delta R_{\gamma\gamma}$ values with the aid of jet substructure analysis~\cite{Ellis:2012sd,Ellis:2012zp,Chakraborty:2017mbz,Wang:2021uyb}, our focus in this work centers on the study of photon-jets with smaller $\Delta R_{\gamma\gamma}$.

It's worth noting that several factors influence the branching ratios of $h_2$ and $a$ decay modes. Firstly, for $h_2\rightarrow\gamma\gamma$, its branching ratio can be significantly enhanced if additional new vector-like charged fermions or other charged scalar/vector particles are introduced. These new particles contribute to loop processes in $h_2\rightarrow\gamma\gamma$, resulting in an elevated branching ratio. The same enhancement applies to $a\rightarrow\gamma\gamma$. Alternatively, if the light pseudoscalar behaves as a gaugphilic ALP, the $a\rightarrow\gamma\gamma$ mode naturally becomes the dominant decay channel. Secondly, for $m_{h_2} > 2m_{\pi}$, the $h_2\rightarrow\pi\pi$ mode becomes the dominant decay channel, with a fraction of $\Gamma (h_2\rightarrow\pi^+\pi^-)/\Gamma (h_2\rightarrow\pi^0\pi^0)\sim 2$~\cite{Gunion:1989we}. However, the light pseudoscalar can exhibit mixing with SM CP-odd mesons if they share the same quantum numbers. For $m_a\lesssim 3m_{\pi}\approx 405$ MeV, the three-body decay modes, $a\rightarrow\pi^0\gamma\gamma$, $2\pi^0\gamma$, $\pi^{+}\pi^{-}\gamma$, are considerably suppressed due to phase space limitations, rendering the branching ratio of $a\rightarrow\gamma\gamma$ nearly equal to one~\cite{Dobrescu:2000jt}. Conversely, for $m_a\gtrsim 0.5$ GeV, isospin-violating decay modes, $a\rightarrow 3\pi$, become accessible, competing with $a\rightarrow\gamma\gamma$. Precise estimation of the decay width is challenging due to QCD uncertainties. However, it's important to note that the branching fractions for $\eta$ meson decays into $\gamma\gamma$, $3\pi^0$, and $\pi^{+}\pi^{-}\pi^0$ are approximately $39.41\%$, $32.68\%$, and $22.92\%$~\cite{Workman:2022ynf}, respectively\footnote{We ignore other decay modes from $\eta$ which are less than $5\%$.}. Since the $\eta$ meson and $a$ share the same quantum numbers, the $a\rightarrow\gamma\gamma$ decay mode remains significant, even for $m_a$ on the order of $1$ GeV.

In an effort to ensure the model independence of our studies, we systematically set the decay branching ratios of $h_2/a\rightarrow\gamma\gamma$, $h_2\rightarrow\pi^0\pi^0$, and $a\rightarrow 3\pi^0$ to unity individually for each analysis. This approach allows our findings to be easily rescaled according to the specific branching ratios associated with each decay channel within any concrete Higgs portal model, thereby enhancing the applicability of our results. Moreover, our focus in this work remains exclusively on the prompt decays of $h_2$ and $a$, deferring the study of long-lived $h_2$ and $a$ to future investigations.

\section{Photon-jet identification with deep learning}
\label{sec:identification}
% {\color{red}
% \begin{itemize}
%    % \item Add the materials from 2203.16703, 
%    % \item Assume the photon-jet signal efficiency to be about $70\%$ and the single photon efficiency to be about $50\%$ (or $20\%$), finally, the $\pi^0$ efficiency to be about $10\%$.
% \end{itemize}
% }

We simulated a lead/liquid-argon (LAr) sampling electromagnetic calorimeter (ECAL) with granularity similar to that of the ATLAS ECAL and a pseudorapidity ($\eta$) coverage of $-0.2 < \eta < 0.2$ using \textsf{GEANT4}~\cite{GEANT4:2002zbu}. It consists of a thin pre-sampling layer and three sampling layers longitudinal in the shower depth.
The first sampling layer is segmented into high-granularity strips in the $\eta$ direction, with a cell size of $0.0031\times 0.098$ in $\Delta \eta \times \Delta \phi$. The pre-sampling layer, second sampling layer and third sampling layer have granularity of $0.025\times 0.1$, $0.025\times 0.0245$ and $0.05\times 0.0245$ in $\Delta \eta \times \Delta \phi$, respectively. Throughout these simulations, we maintain a fixed value of zero for $\eta$.

We recorded the energy deposits of the photon-jets---which arise from the decay of $h_2$ or $a$, the single photon, or $\pi^0\rightarrow \gamma \gamma$ backgrounds---in each cell of the calorimeter.
Our study focuses on four benchmark masses of $h_2$ and $a$; specifically, 0.45 GeV, 0.6 GeV, 0.8 GeV, and 1 GeV. For each process, we generated a sample of 100,000 events with the energy of the source particle uniformly distributed between 40 and 250 GeV. For each sample, $70\%$ of the events were used as the training set and the rest for the test set.

The three subsections that follow describe the our implementations of the Boosted Decision Tree (BDT), Convolutional Neural Network (CNN), and Particle Flow Network (PFN). Our full code is available at \url{https://github.com/womogenes/photon-jet}. For each model class (BDT, CNN, PFN), we trained three distinct models for the $h_2\rightarrow\pi^0\pi^0$, $a\rightarrow\gamma\gamma$, and $a\rightarrow3\pi^0$ tasks.

\subsection{BDT implementation}
We employed Gradient BDTs~\cite{Friedman:2001wbq} with multi-class classification as our first method for separating signal and background processes. BDTs are composed of an ensemble of decision trees that iteratively build on each other to generate outputs more accurate than each tree can produce on its own.

At a high level, training a Gradient BDT involves performing a number of ``boosting iterations," where each iteration adds a new decision tree to the ensemble. These new trees are intended to remediate mistakes made by earlier trees, and have their initial parameters tuned by gradient descent. More details can be found in Ref.~\cite{Friedman:2001wbq}.

\begin{table}[!ht]
   % \centering
   \setlength\extrarowheight{2pt} 
    \begin{tabularx}{\textwidth}{c|X|c}
        \hline
        Name           & Definition         &Relevant ECAL layer                                                    \\
        \hline
        $f_1$          & Fraction of energy in the first layer                          &\multirow{6}{*}{First layer}           \\
        $f_\mathrm{side}$     & Fraction of energy outside core of three central strips, but within seven strips            \\
        $w_{s3}$        & Lateral shower width, $\sqrt{\frac{\Sigma E_i(i-i_{\text{max}})^2}{\Sigma E_i}}$, $i$ runs over three strips around maximum strip                        \\
        $w_{s20}$      & Lateral shower width, $\sqrt{\frac{\Sigma E_i(i-i_{\text{max}})^2}{\Sigma E_i}}$, $i$ runs over 20 strips around maximum strip                           \\ 
        $\Delta E_{s}$ & Difference between the energy associated with the
        second maximum, and the energy reconstructed in the strip with the minimal value found
        between the first and second maxima \\
        $E_\mathrm{ratio}$    & Ratio of the energy difference associated with the
        largest and second largest energy deposits over the
        sum of these energies \\
        \hline
        $R_{\eta}$     & Ratio in $\eta$ of cell energies in $3 \times 7$ versus $7 \times 7$ cell, both centered around the maxima &\multirow{3}{*}{Second layer}\\
        $R_{\phi}$     & Ratio in $\phi$ of cell energies in $3\times3$ and $3\times7$ cells, both centered around the maxima       \\
        $w_{\eta2}$ &Lateral shower width, $\sqrt{\frac{\Sigma E_i \eta_i^2}{\Sigma E_i} - (\frac{\Sigma E_i \eta_i}{\Sigma E_i})^2}$, $i$ runs over the cells within $3 \times5 $ window around the maxima \\
        \hline
    \end{tabularx}
    \caption{The shower shape variables used in the BDT studies. }
    \label{tab:bdt_vars}
\end{table}

\begin{figure}
    \begin{center}
        \subfloat[]{\includegraphics[width=.33\columnwidth]{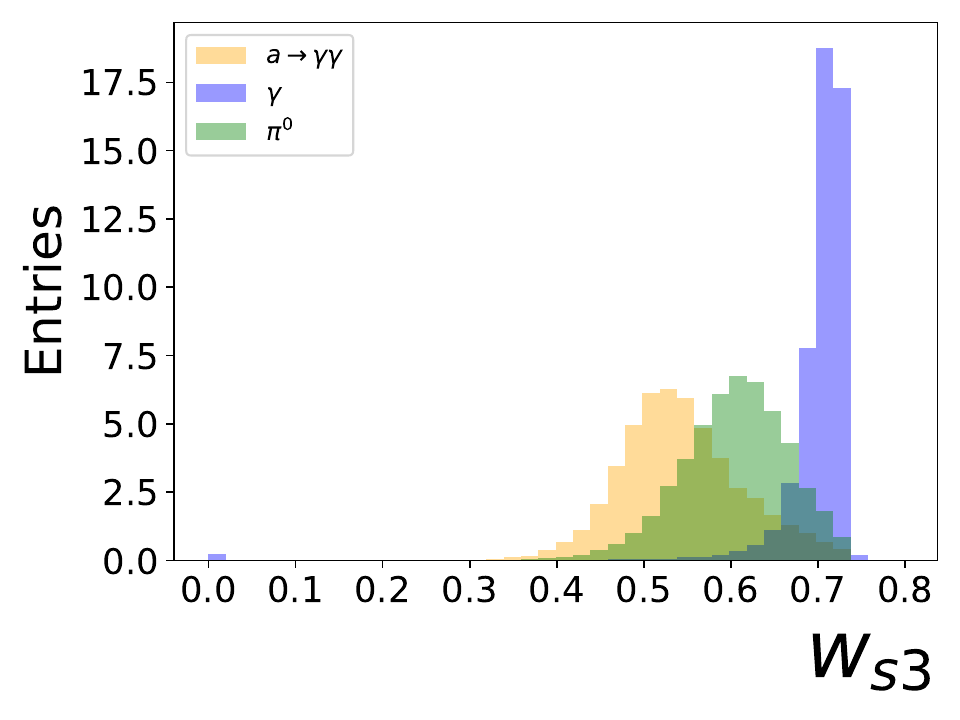}}
        \subfloat[]{\includegraphics[width=.33\columnwidth]{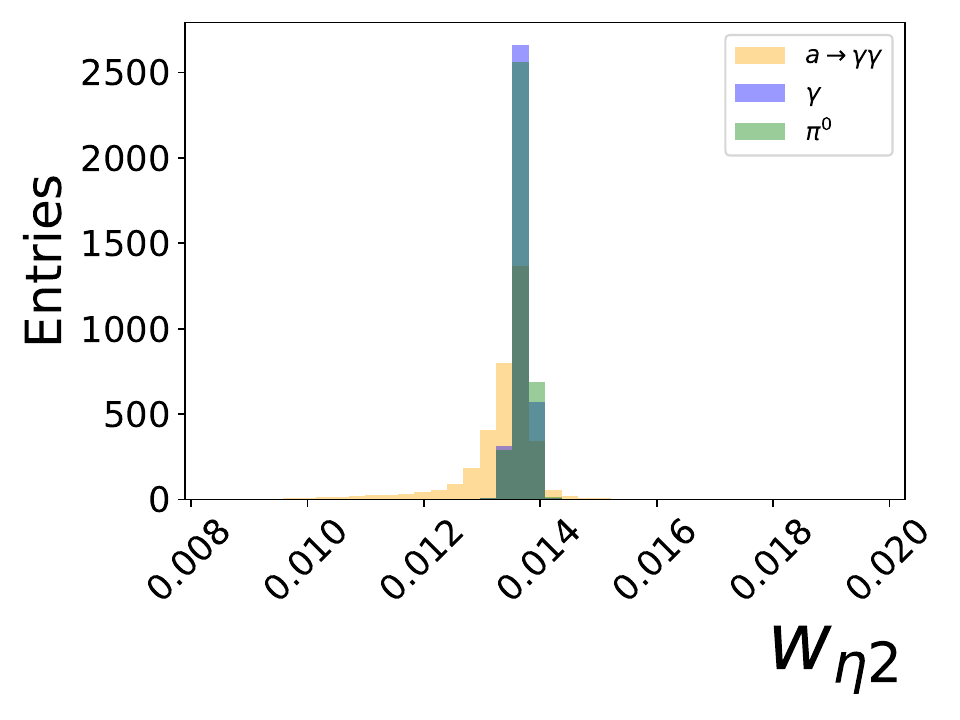}}
        \subfloat[]{\includegraphics[width=.33\columnwidth]{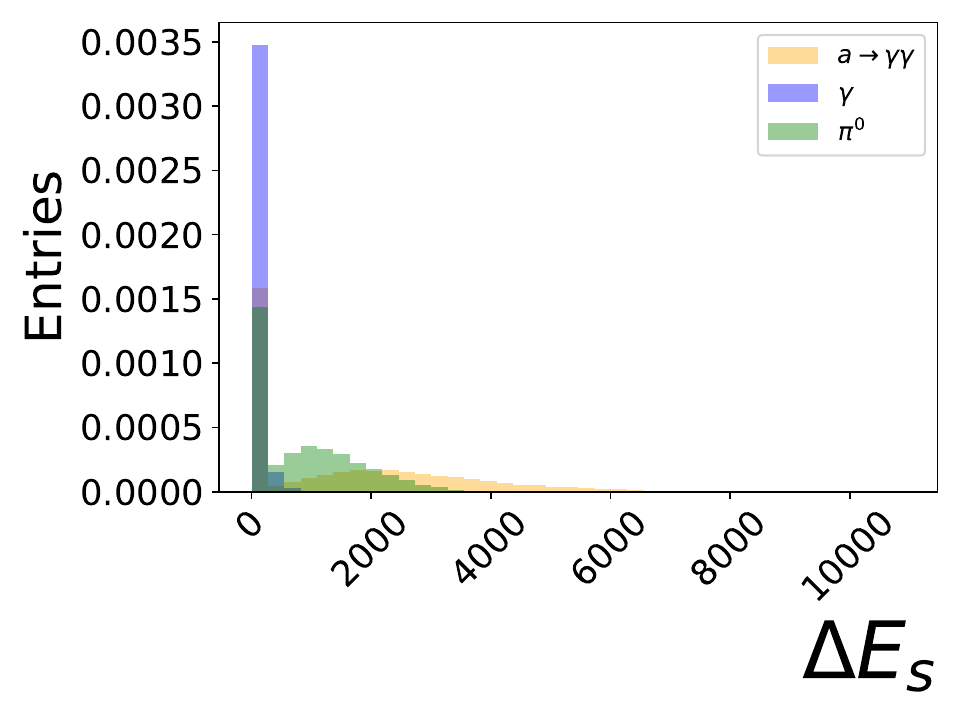}} \\
        
        \subfloat[]{\includegraphics[width=.33\columnwidth]{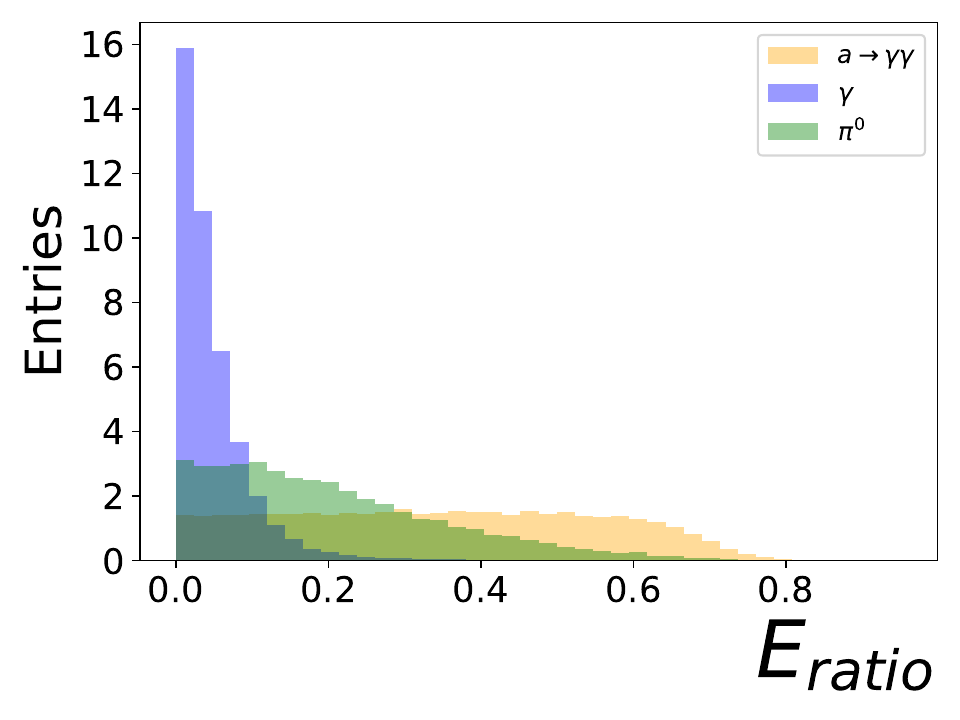}}
        \subfloat[]{\includegraphics[width=.33\columnwidth]{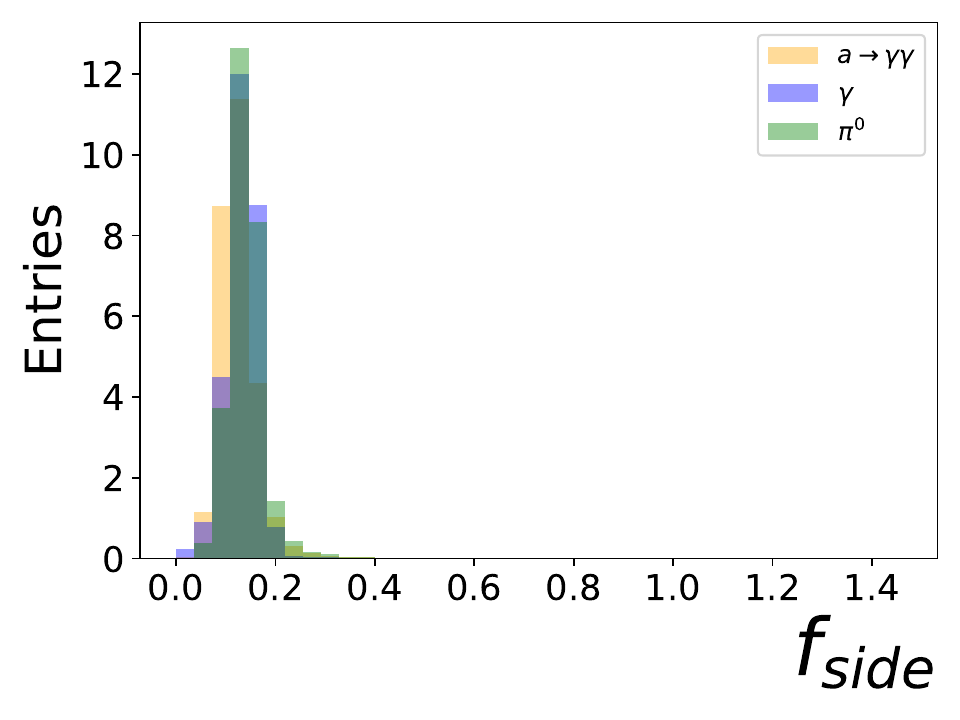}}
        \subfloat[]{\includegraphics[width=.33\columnwidth]{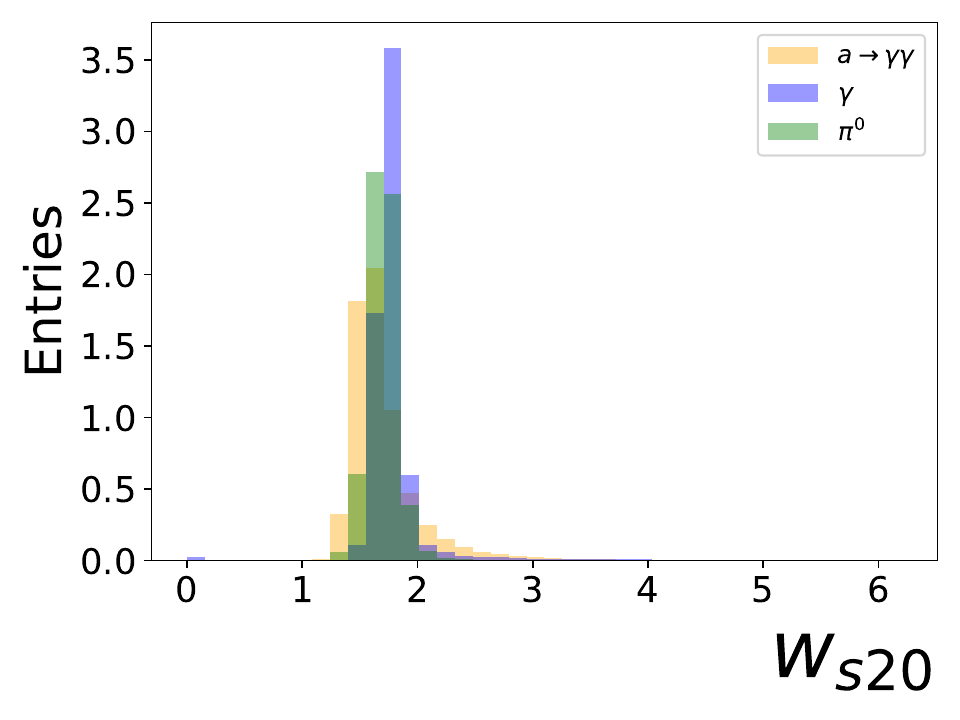}} \\

        \subfloat[]{\includegraphics[width=.33\columnwidth]{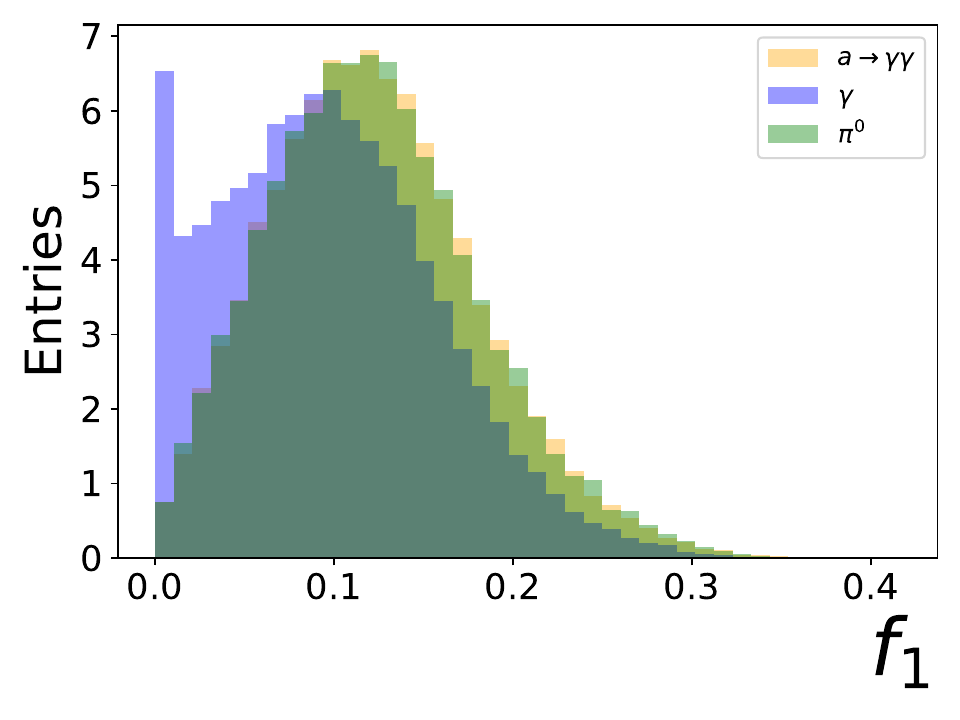}}
        \subfloat[]{\includegraphics[width=.33\columnwidth]{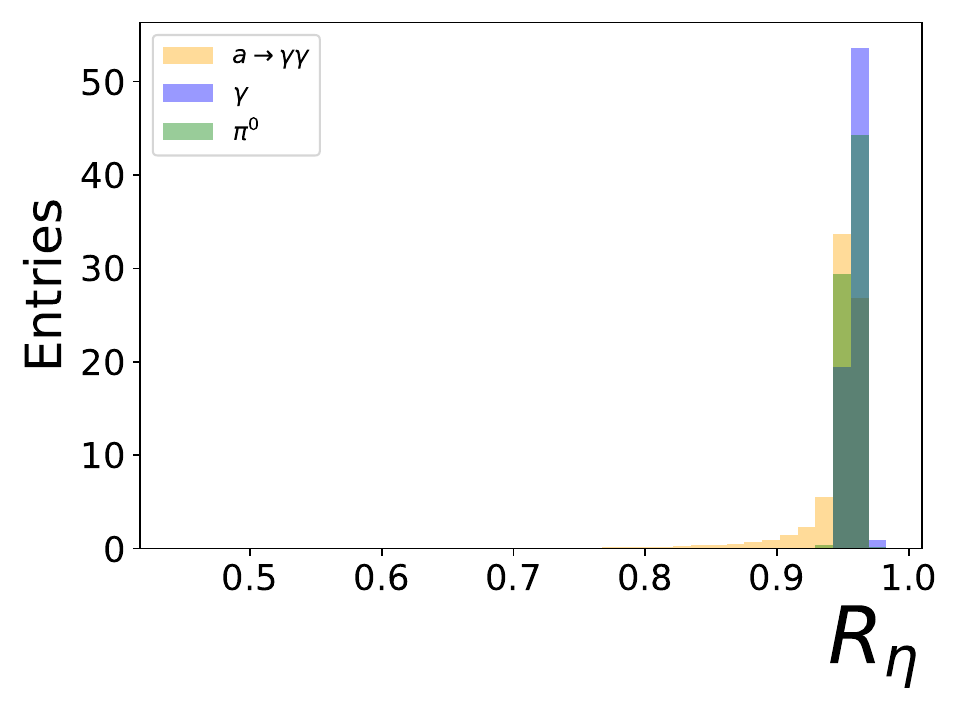}}
        \subfloat[]{\includegraphics[width=.33\columnwidth]{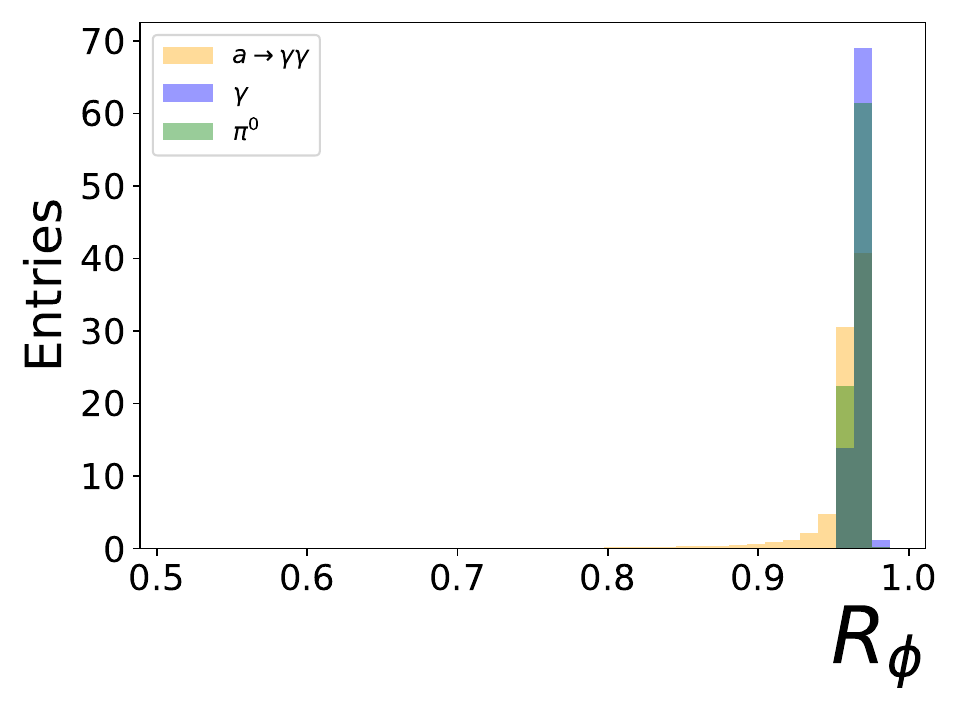}} \\
    \end{center}
    \caption{
        \label{fig:bdt_var_hists}
        The BDT variables of photon-jet from $a\rightarrow \gamma \gamma$ ($m_a = 1$ GeV) and relevant backgrounds from the single photon and the neutral pion. The $a$, $\gamma$ and $\pi^0$ have energy in the range of $[40, 250]$ GeV.
    }
\end{figure}

The features we used as inputs to our BDTs are the shower shape variables $f_1$,  $f_\mathrm{side}$, $w_{s3}$, $w_{s20}$, $\Delta E_{s}$, $E_\mathrm{ratio}$, $R_{\eta}$, $R_{\phi}$, $w_{\eta2}$, as used in Ref.~\cite{ATLAS:2018fzd}. The definitions of these variables are provided in Table~\ref{tab:bdt_vars}, and their distributions are visualized in Fig.~\ref{fig:bdt_var_hists}.

%% The BDT is trained in four bins of energy: $[40, 100]$ GeV, $[100, 150]$ GeV, $[150, 200]$ GeV and $[200, 250]$ GeV.
%% What exactly does it mean to be "trained in four bins of energy"?
We used \textit{sklearn.ensemble.GradientBoostingClassifier} from the scikit-learn Python package~\cite{Pedregosa:2011ork} to implement our GBDTs. The most important hyperparameters are described below:
\begin{enumerate}
    \item The depth of each decision tree is limited to a maximum of 5. We observed that any higher values tended to result in overfitting of training data.
    \item The number of boosting stages is set to $100$, because after that we observed very minimal performance gains.
    \item The learning rate is set to $0.5$, which means that after every boosting stage, the contribution of the added decisions tree gets cut in half. Lower values tended to result in underfitting or slower training while higher values tended to result in overfitting.
\end{enumerate}

\subsection{CNN implementation}

\begin{figure}
    \begin{center}
        \subfloat[]{\includegraphics[width=.33\columnwidth]{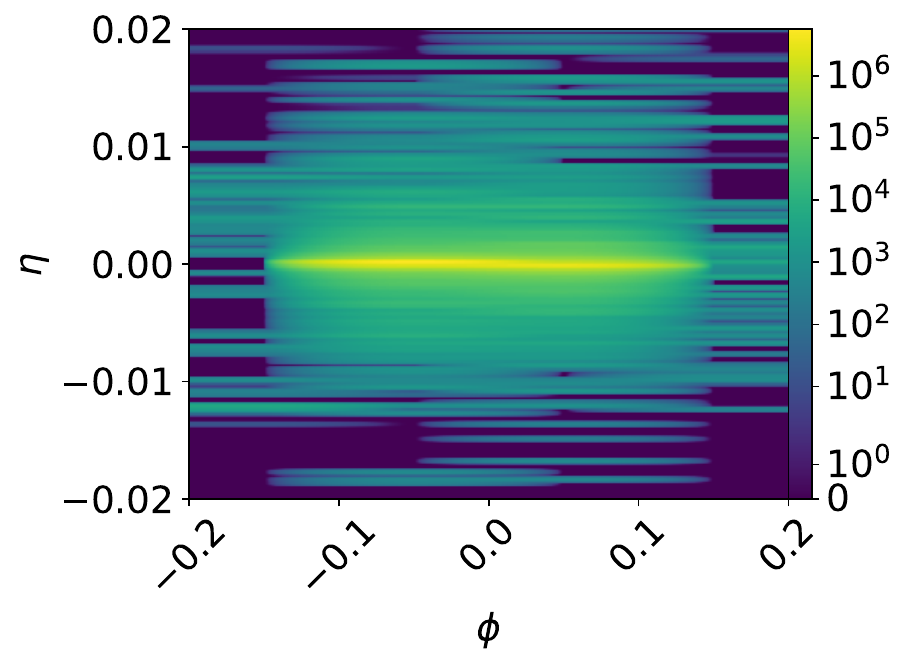}}
        \subfloat[]{\includegraphics[width=.33\columnwidth]{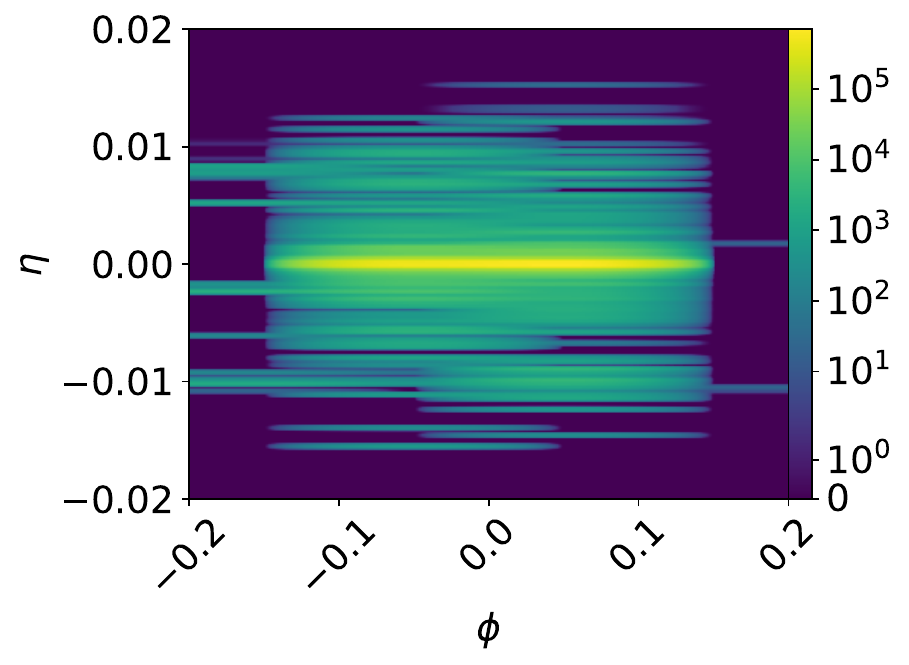}}
        \subfloat[]{\includegraphics[width=.33\columnwidth]{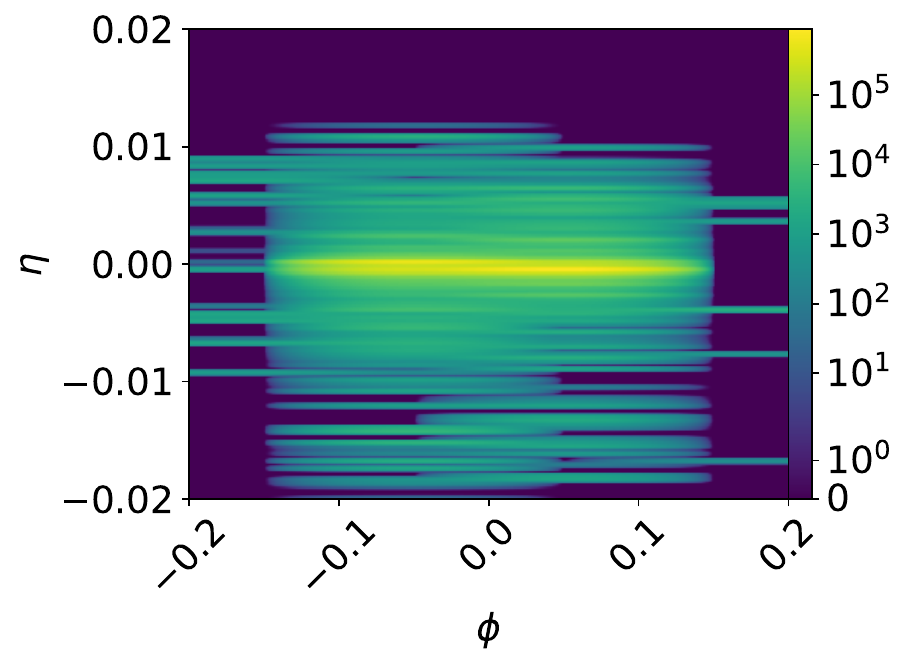}}
    \end{center}
    \caption{
        \label{fig:ene_deposits}
        The deposited energy per cell of (a) photon-jet from $a\rightarrow \gamma \gamma$ ($m_a = 1$ GeV) (b) $\gamma$ (c) $\pi^0$ at the first layer of the ECAL. The $a$, $\gamma$ and $\pi^0$ have energy in the range of $[40, 250]$ GeV.
    }
\end{figure}

We show the energy deposits for all cells per ECAL layer represented as a 2D image of dimension $N_{\text{cells}} (\phi) \times N_{\text{cells}} (\eta)$ in Fig.~\ref{fig:ene_deposits}, where $N_{\text{cells}} (\phi)$ and $N_{\text{cells}} (\eta)$ are the number of cells in $\phi$ and $\eta$ direction, respectively. The value for each cell represents the energy deposited in it. These images are used as the inputs to our CNNs.

\begin{figure}
    \centering
    \input{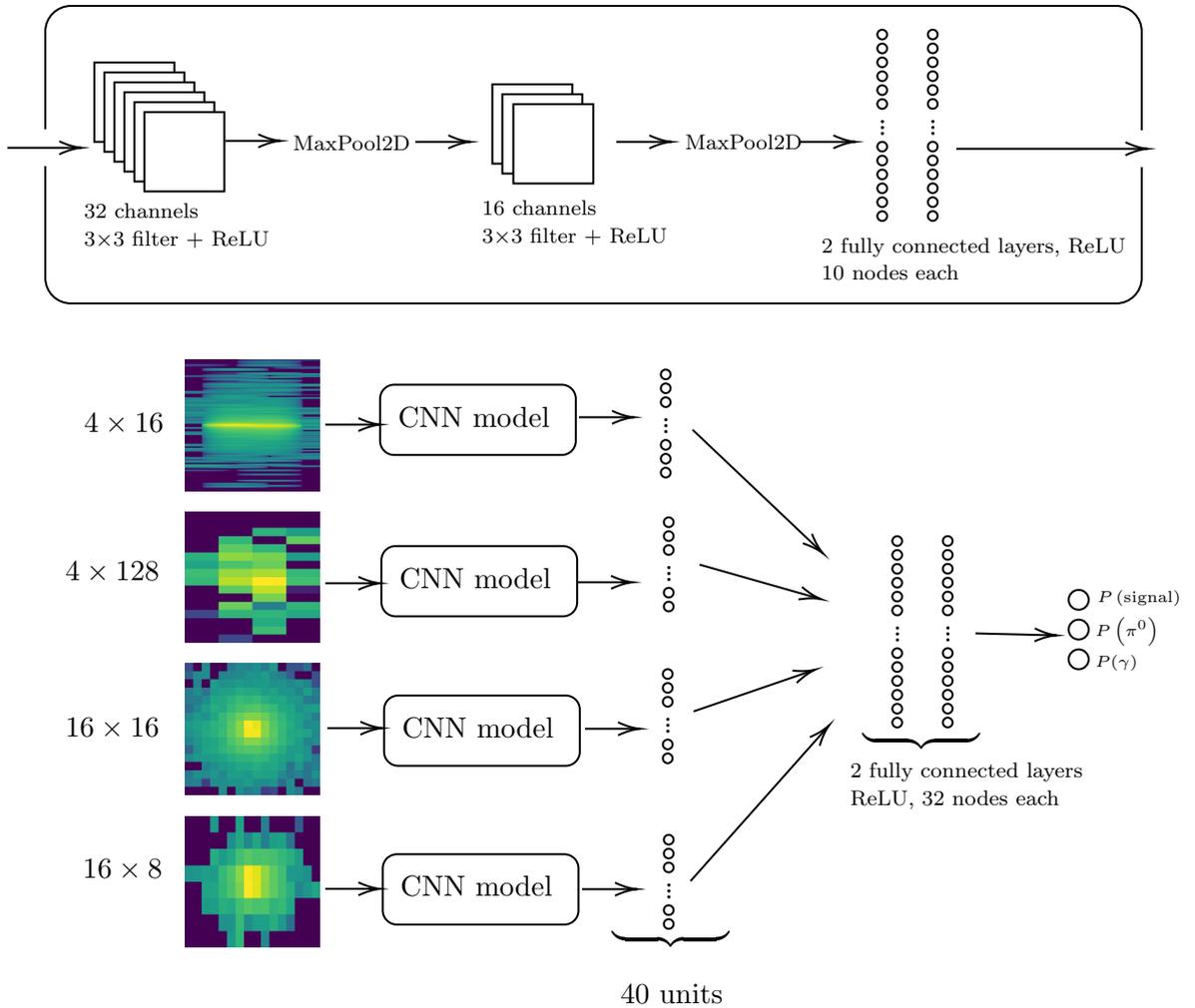}
    \caption{Diagram of the CNN architecture.}
\label{fig:CNN_architecture_diagram}
\end{figure}

An illustration of the CNN architecture is shown in Fig.~\ref{fig:CNN_architecture_diagram}. Four separate CNN models exist: one for each of the four ECAL layers. Each CNN model is constructed with two convolutional layers with filters of size $3\times 3$ and stride 1, with the \textit{rectified linear unit} (ReLU)~\cite{Agarap:2018uiz} as its activation function. Each convolutional layer is followed by a max-pooling layer of size $2\times 2$. A flatten layer is used to convert the 2D output array from the pooling layer to 1D array. Finally, the 1D arrays from the four CNN models are concatenated before being fed to two fully connected layers with 32 nodes and the ReLU activation. The final output layer is a fully connected layer with the \textit{softmax} activation function and number of nodes equal to the number of distinct classes providing a multi-class output.
We used the \textsf{Tensorflow}~\cite{Abadi:2016kic} and \textsf{Keras} packages to implement our CNNs. For training, we utilized the categorical \textit{cross-entropy} loss function and the \textit{Adam} optimizer~\cite{Kingma:2014vow}.

%% TODO: Add figure showing architecture of the PFN (William)
\subsection{PFN implementation}

In addition to the BDT and CNN classification methods, we employed PFNs to identify the photon-jets of our signal processes. Developed by Komiske et al.~\cite{Komiske:2018cqr}, PFNs are a class of deep learning model that take as input a jet represented in point-cloud form (i.e. an unordered set of feature vectors) and output a vector of probabilities for classification.
%% We used PFNs to separate the photon-jets of each of the three signal processes ($h_2 \rightarrow \pi^0\pi^0$, $a\rightarrow \gamma\gamma$, and $a\rightarrow 3\pi^0$) from those of the background (photons and neutral pions). 
%% Unneeded ^
Komiske et al. demonstrated that PFNs are able to model \textit{any permutation-invariant} function on a point cloud. That is to say, the order in which points are stored in the point cloud does not impact the output of a PFN, and any function with this property can be represented as a PFN\footnote{In Ref.~\cite{Komiske:2018cqr}, Komiske et al. formulate a close relative of the PFN called the Energy Flow Network (EFN). We considered using EFNs in our studies, but found that they are less performant than PFNs because the former is essentially a sub-case of the latter and less generalizable.}.

PFNs cannot directly take our ECAL images as input, so we must first ``devoxelize" the images into point clouds. To accomplish this, the array of each sampling layer is transformed into an unordered set of ``points" (feature vectors), one for each cell in the sampling layer. Each point is a four-tuple $(\eta, \phi, E, \ell)$ where:
\begin{enumerate}
    \item $\eta$ and $\phi$ are the original coordinates of the cell (normalized to have mean zero and standard deviation 1),
    \item $E$ is the energy of the cell in GeV, and
    \item $\ell$ is the index of the layer (ranging from 1 to 4 for the pre-sampling to third sampling layers).
\end{enumerate}
This creates an unordered set of $960$ feature vectors (particles) for each jet. As a final step, we filter out all cells with zero energy, as those should have no effect on the model's output and we can take greater advantage of the variable-length nature of PFN inputs. This also helps compress data for faster training and lower overfitting.

Once a jet has been devoxelized into a point cloud $\mathcal{C}$, it is classified using the equation
\begin{equation}
    \label{Eq:PFN}
    \text{PFN}(\mathcal{C}) = F\!\left(\sum_{\vec{p}\in\mathcal{C}} \Phi(\vec p)\right)\!,
\end{equation}
where $F$ and $\Phi$ are vector functions approximated by  deep neural networks (DNNs). Essentially, every particle in $\mathcal{C}$ is a four-dimensional vector which gets mapped to a 128-dimensional ``latent space" using the function $\Phi: \mathbb{R}^4\rightarrow\mathbb{R}^{128}$. The latent mappings of each particle are then summed up, which we implement using a \textit{tf.math.reduce\_sum} layer in Tensorflow, and this 128-dimensional vector sum is then converted to final classification probabilities using the function $F: \mathbb{R}^{128}\rightarrow\mathbb{R}^3$.

\begin{figure}
    \begin{center}
    \input{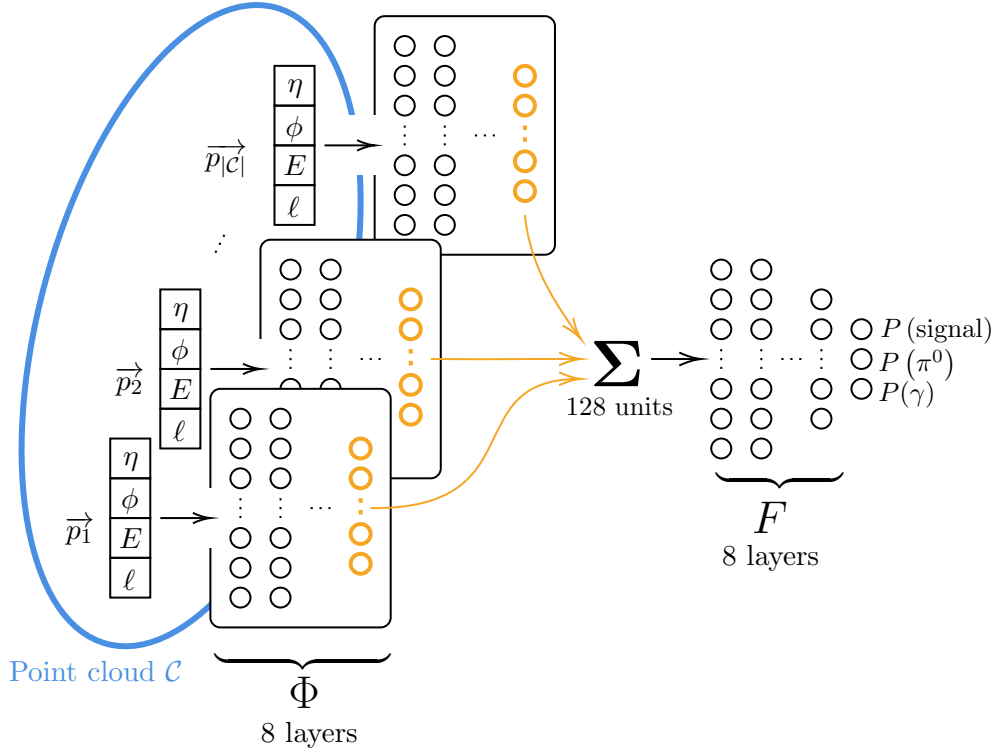}
    \end{center}
    \caption{Diagram of the PFN architecture.}
    \label{fig:PFN_architecture_diagram}
\end{figure}

%% Explain that I chose this architecture after hyperparameter search
%% Do more hyperparameter tuning
%% Ask why we chose the particular 8 features for the BDT
%% Explain how latent space correlates to high-level features
For our models, both $\Phi$ and $F$ are represented as DNNs with 8 hidden layers of sizes $(256, 256, 256, 256, 128, 128, 128, 128)$, with $\Phi$ having an additional input layer of size 4 (the number of features per particle) and $F$ having an additional output layer of size 3 (the number of classes). The ReLU activation function~\cite{Agarap:2018uiz} is used in all layers except the output layer of $F$, which uses \textit{softmax}. A diagram of our PFN's architecture is shown in Fig.~\ref{fig:PFN_architecture_diagram}. 

The \textit{categorical cross-entropy} loss function and \textit{adam} optimizer~\cite{Kingma:2014vow} are used to train the weights of $F$ and $\Phi$. We implemented our model and training using the TensorFlow \textit{Keras} library~\cite{Abadi:2016kic}.

\subsection{Performance}

\begin{figure}
    \begin{center}
        \subfloat[]{\includegraphics[width=.33\columnwidth]{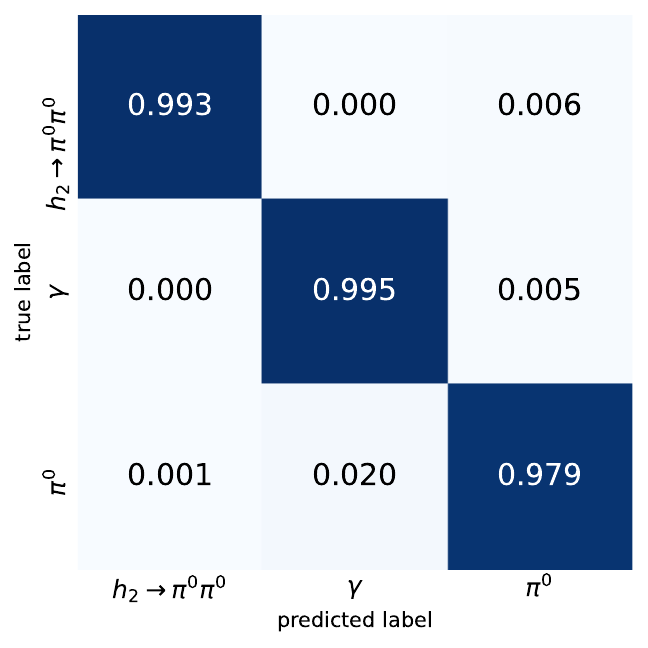}}
        \subfloat[]{\includegraphics[width=.33\columnwidth]{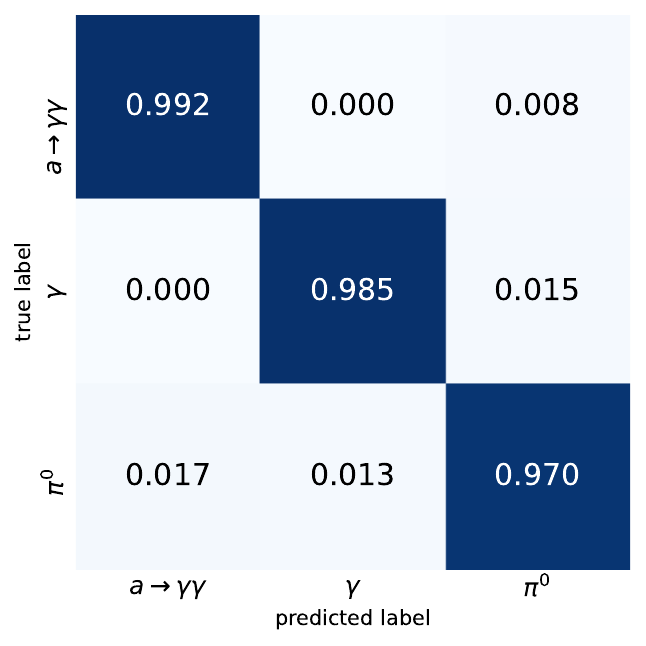}}
        \subfloat[]{\includegraphics[width=.33\columnwidth]{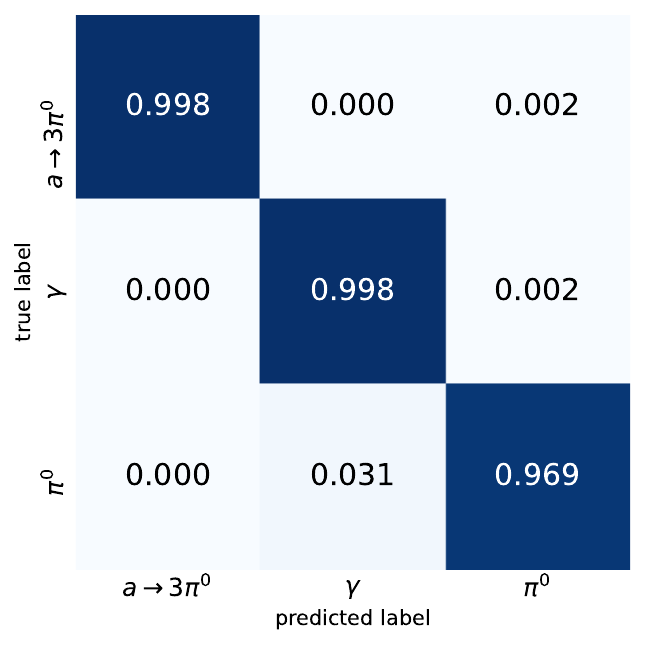}}
    \end{center}
    \caption{
        \label{fig:CNN_confmatrix}The normalized confusion matrix for distinguishing between the photon-jet produced in the process of (a) $h_2 \rightarrow \pi^{0} \pi^{0}$ (b) $a\rightarrow \gamma \gamma$ (c) $a\rightarrow 3\pi^{0}$, and the single photon and $\pi^0$ backgrounds for the test set using CNN. The mass of $h_2$ and $a$ is assumed to be 1~GeV.
    }
\end{figure}

Fig.~\ref{fig:CNN_confmatrix} shows the confusion matrix for distinguishing between photon-jets produced in three different signal processes and the same $\gamma$, $\pi^0$ backgrounds with the test set using CNN. The overall identification efficiency of the photon-jet is above 99.2\% with $\gamma$ rejection rate above 99.9\% and $\pi^0$ rejection rate above 99.8\%.

\begin{figure}
    \begin{center}
    \subfloat[]{\includegraphics[width=.33\columnwidth]{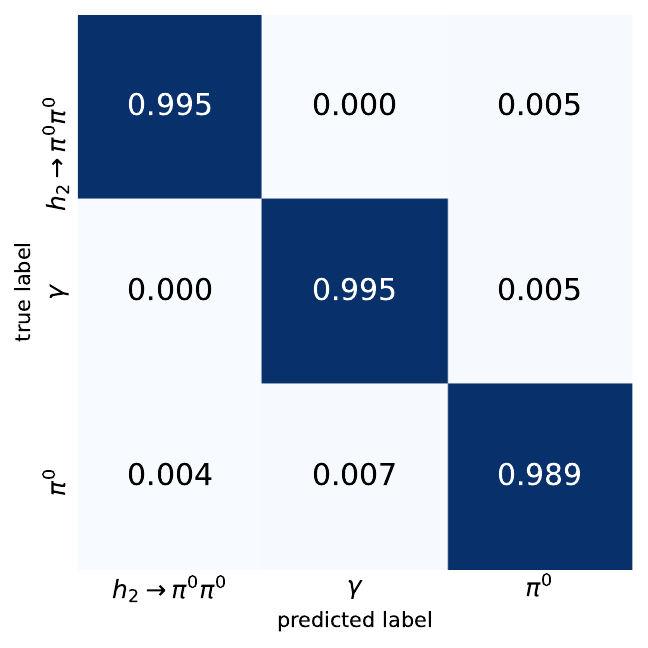}}
    \subfloat[]{\includegraphics[width=.33\columnwidth]{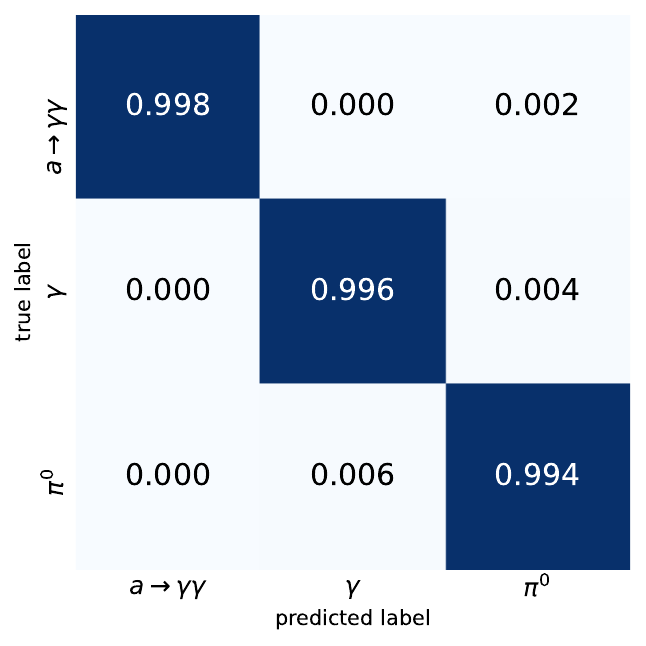}}
    \subfloat[]{\includegraphics[width=.33\columnwidth]{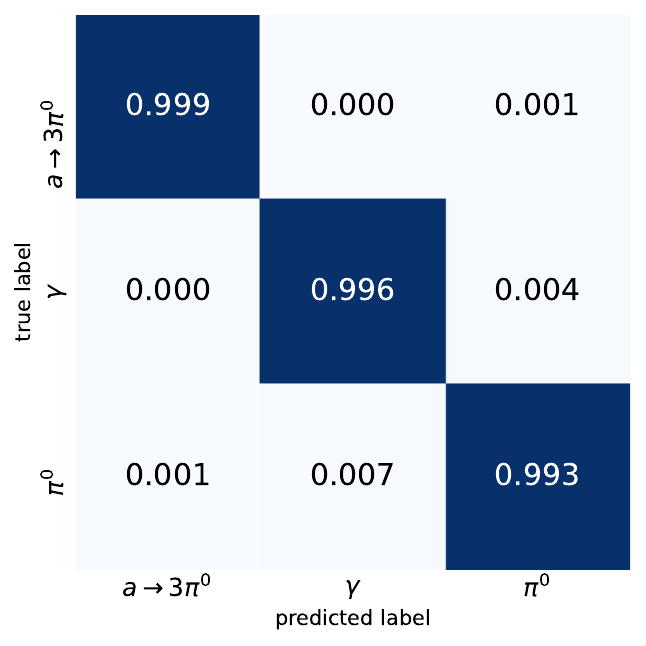}}
    \end{center}
    \caption{
        \label{fig:PFN_confmatrix}
        Analogous confusion matrices to those in Fig.~\ref{fig:CNN_confmatrix} for PFNs instead of CNNs.
        %The average mass of $h_2$ and $a$ is also assumed to be 1~GeV. 
    }
\end{figure}

 An analogous confusion matrix for our PFNs is shown in Fig.~\ref{fig:PFN_confmatrix}. They outperform our CNNs in signal identification efficiency but are behind in background rejection rates. The average identification efficiency exceeds 99.7\% (99.5\% for the $h_2 \rightarrow\pi^0\pi^0$ task, 99.8\% for the $a\rightarrow\gamma\gamma$ task, and 99.9\% for the $a\rightarrow 3\pi^0$ task), while its overall background rejection rate exceeds 99.8\% for the $\pi^0$ background and 99.9\% for the $\gamma$.

%% Table~\ref{tab:eff_compare} shows that BDTs vastly underperform CNNs and PFNs, suggesting that only more sophisticated deep learning approaches are capable of accurately interpreting the complexity of photon-jets.

\begin{figure}
    \centering
    \includegraphics[width=0.7\textwidth]{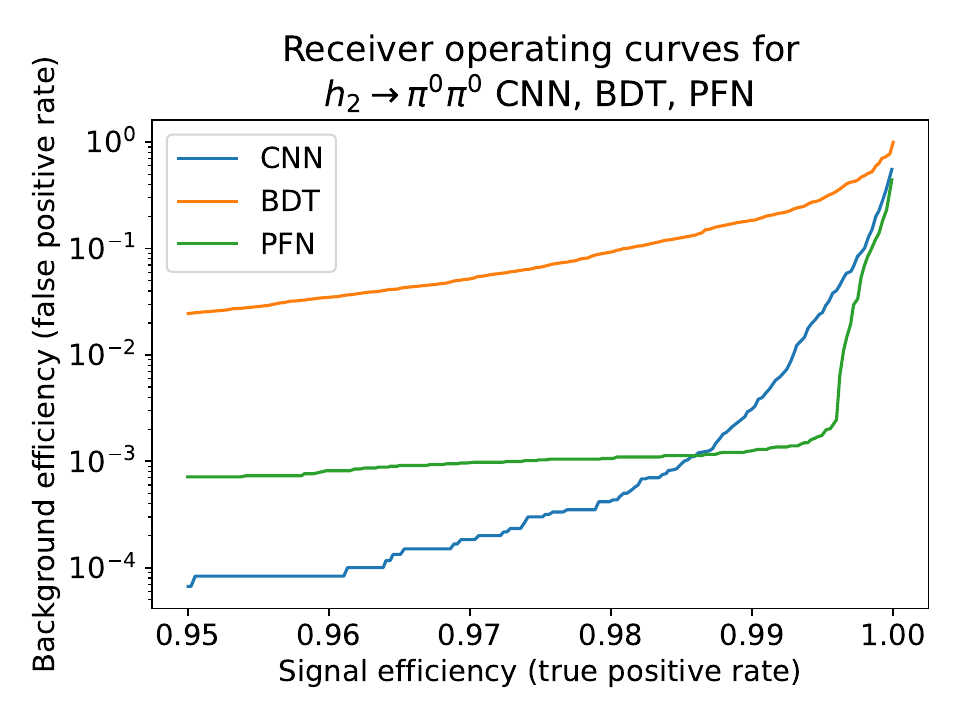}
    \caption{Receiver operating characteristic (ROC) curve for all three model types on the $h_2\rightarrow\pi^0\pi^0$ identification task.}
    \label{fig:roc_curves}
\end{figure}

\begin{table}[]
    \centering
    \begin{tabular}{lllrr}
    \hline
     Model                            & $\text{Rej}_{90\%}$   & $\text{Rej}_{95\%}$   &   $\text{Rej}_{99\%}$ &   $\text{Rej}_{99.5\%}$ \\
    \hline
     CNN ($h_2\rightarrow\pi^0\pi^0$) & 19968                 & 14976                 &                   327 &                      39 \\
     BDT ($h_2\rightarrow\pi^0\pi^0$) & 144                   & 41                    &                     5 &                       3 \\
     PFN ($h_2\rightarrow\pi^0\pi^0$) & 2406                  & 1399                  &                   802 &                     573 \\\hline
     CNN ($a\rightarrow\gamma\gamma$) & 19944                 & 3324                  &                    20 &                      11 \\
     BDT ($a\rightarrow\gamma\gamma$) & 188                   & 58                    &                     7 &                       4 \\
     PFN ($a\rightarrow\gamma\gamma$) & N/A                   & N/A                   &                  1200 &                     484 \\\hline
     CNN ($a\rightarrow3\pi^0$)       & N/A                   & 59885                 &                  1619 &                     166 \\
     BDT ($a\rightarrow3\pi^0$)       & 253                   & 90                    &                    10 &                       5 \\
     PFN ($a\rightarrow3\pi^0$)       & N/A                   & N/A                   &                  1253 &                     676 \\
    \hline
    \end{tabular}
    \caption{Comparison of model efficiency. For each of our nine models (CNN, BDT, and PFN trained across three tasks), we evaluated the rejection rates (defined as 1 over background efficiency) at 90\%, 95\%, 99\%, and 99.5\% working points. The rejection rate appears as N/A if a model performs so well that no threshold makes it dip below the working point.}
    \label{tab:eff_compare}
\end{table}

%% Add the BDT (William)
%% Make a note earlier on that we don't need BDT confusion matrix
\begin{figure}
    \begin{center}
        \subfloat[]{\includegraphics[width=.33\columnwidth]{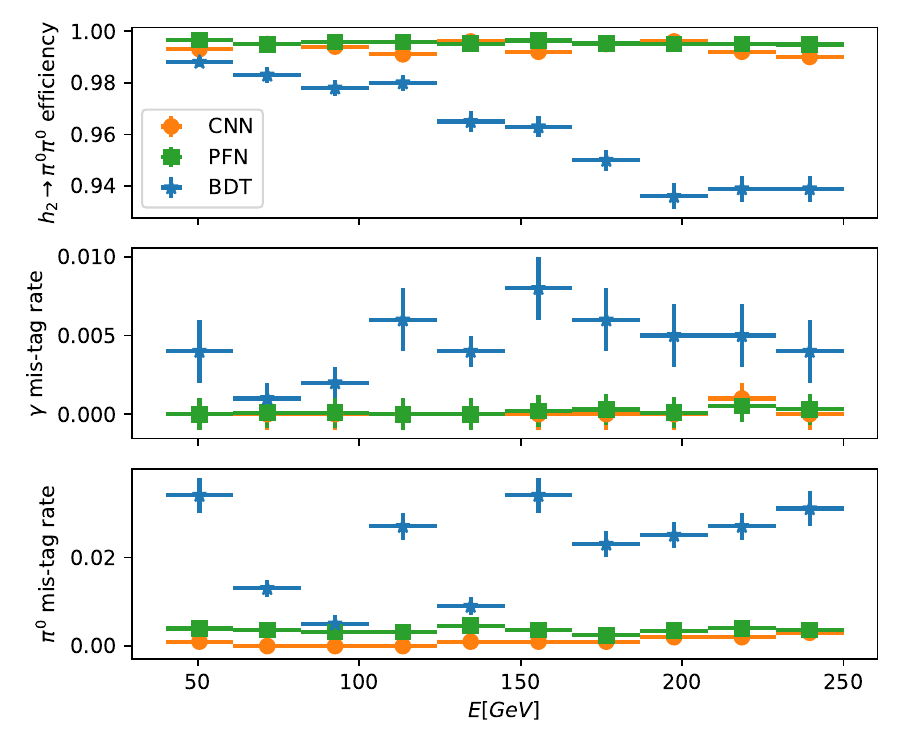}}
        \subfloat[]{\includegraphics[width=.33\columnwidth]{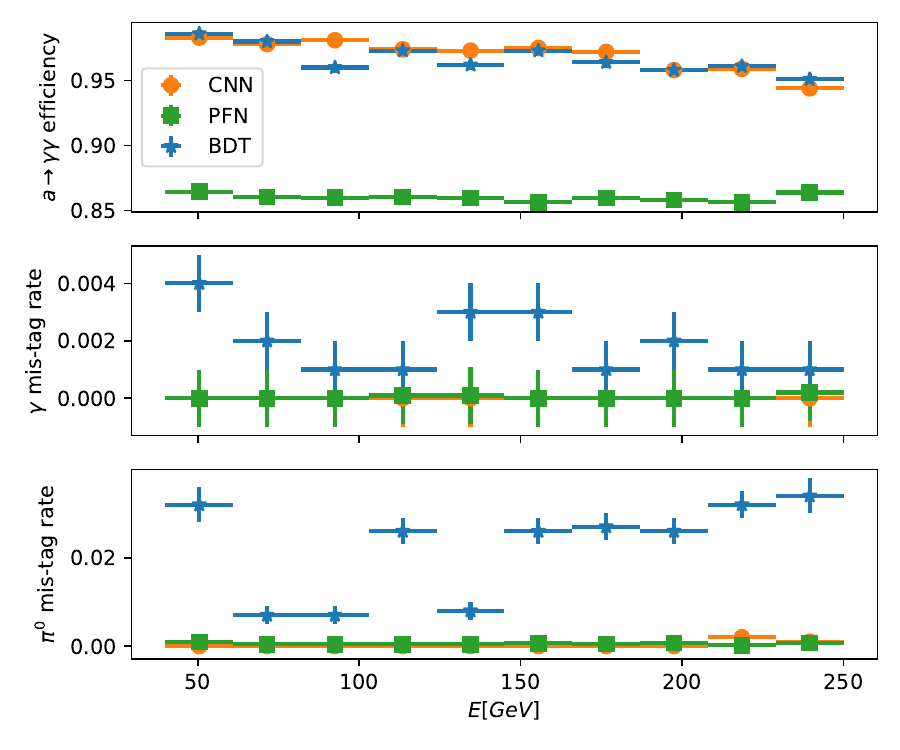}}
        \subfloat[]{\includegraphics[width=.33\columnwidth]{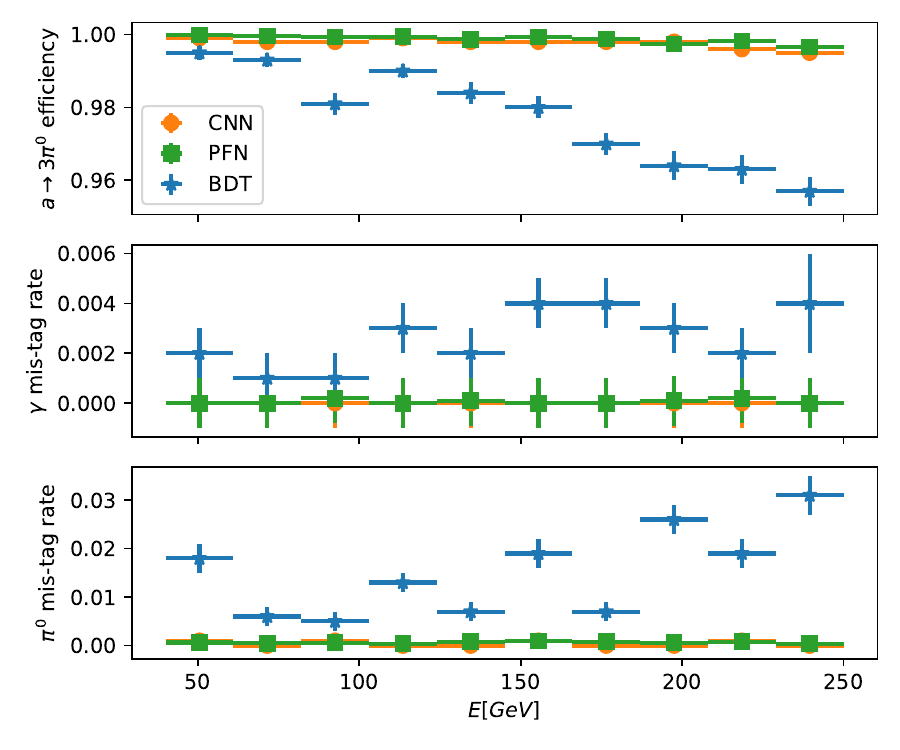}}
    \end{center}
    \caption{
        \label{fig:PFN_CNN_comp_eff}
        Comparison of identification efficiency of photon-jets between CNNs (orange), PFNs (green), and BDTs (blue) in the process of (a) $h_2\rightarrow\pi^0\pi^0$ (b) $a\rightarrow\gamma\gamma$ (c) $a\rightarrow 3\pi^0$. The top row represent signal signatures and the bottom two rows show background signatures. The $y$-coordinate of each point represents the proportion of events with a given energy ($x$-axis) that are classified as signal, with associated error bars.
    }
\end{figure}

Fig.~\ref{fig:roc_curves}, Table~\ref{tab:eff_compare}, and Fig.~\ref{fig:PFN_CNN_comp_eff} compare the efficiency and mis-tag rates of our CNNs, BDTs, and PFNs across different working points and energy bins. (``Efficiency" refers to the proportion of events tagged as signal, and ``mis-tag rate" refers to the proportion of background events tagged as signal.)

Our results indicate that BDTs are not well-suited to the photon-jet tasks, as its signal efficiency remains consistently below that of CNNs and PFNs across all energy bins. This could be explained by the fact that BDTs are less structurally complex than other deep learning models, and therefore are incapable of modeling the finer patterns present in photon-jets.
%% Insert reference?

We observe that the CNNs have significantly higher efficiency on the $h_2 \rightarrow \pi^{0} \pi^{0}$ tagging task than $a\rightarrow \gamma \gamma$ and $a\rightarrow 3\pi^{0}$ tasks. Also, PFNs have better signal efficiencies than CNNs, especially for the $a\rightarrow \gamma \gamma$ task. This lies outside the statistically significant range of the error bars. However, PFNs have worse mis-tag rates than CNNs, and this is especially pronounced for the $h_2 \rightarrow \pi^{0} \pi^{0}$ task. We will use the analysis results from our CNNs in Sec.~\ref{sec:results}, though PFNs achieve similar levels of performance and could be another suitable choice.

\subsection{Interpretation of PFNs}

We now seek to establish the interpretability of our PFNs, in hopes of elucidating why it performs so well.

%% Insert 3 tables, one for each model
\begin{table}[ht]
    \renewcommand{\arraystretch}{1.2}
    \centering
    \begin{tabularx}{\textwidth}{>{\raggedright\arraybackslash\setlength{\baselineskip}{0.9\baselineskip}}X|l|l|l}
        \hline
         BDT variable & $h_2 \rightarrow \pi^{0} \pi^{0}$ PFN & $a\rightarrow \gamma \gamma$ PFN & $a\rightarrow 3\pi^0$ PFN \\
        \hline
         \texttt{depth\_weighted\_total\_e}: Summed energy across all 960 calorimeter cells, directly weighted by layer (0 for pre-sampling layer, 1 for the first, 2 for the second, 3 for the third) & $\Sigma_{117}$ (0.998) & $\Sigma_{93}$ (0.998) & $\Sigma_{73}$ (0.999) \\
         \texttt{total\_e}: Summed energy across all 960 calorimeter cells, unweighted & $\Sigma_{88}$ (0.999) & $\Sigma_{93}$ (0.997) & $\Sigma_{19}$ (0.996) \\
         \texttt{depth\_weighted\_total\_e2}: Summed energy squared across all cells, directly weighted by layer & $\Sigma_{68}$ (0.998) & $\Sigma_{93}$ (0.996) & $\Sigma_{73}$ (0.997) \\
         \texttt{prelayer\_e}: Summed energy across pre-sampling layer & $\Sigma_{56}$ (0.847) & $\Sigma_{7}$ (0.841) & $\Sigma_{126}$ (0.876) \\
         \texttt{firstlayer\_x2}: Summed energy across first layer, weighted by $x$-coordinate squared $(\phi^2)$ & $\Sigma_{41}$ (0.834) & $\Sigma_{78}$ (0.812) & $\Sigma_{96}$ (0.907) \\
         \texttt{firstlayer\_y2}: Summed energy across first layer, weighted by $y$-coordinate squared $(\eta^2)$ & $\Sigma_{12}$ (0.820) & $\Sigma_{57}$ (0.722) & $\Sigma_{121}$ (0.887) \\
        \hline
    \end{tabularx}
    \caption{\label{tab:PFN_feature_interp} Physical interpretation of high-level features in PFN models. Pearson correlation coefficients are given between six BDT features and their most highly correlated $\Sigma$~units in each PFN. Only the six most prominent BDT features are shown.}
\end{table}

%% Just keep the 1 figure
%% Leave some figures in the appendix
\begin{figure}
    \centering
    \includegraphics[width=\linewidth]{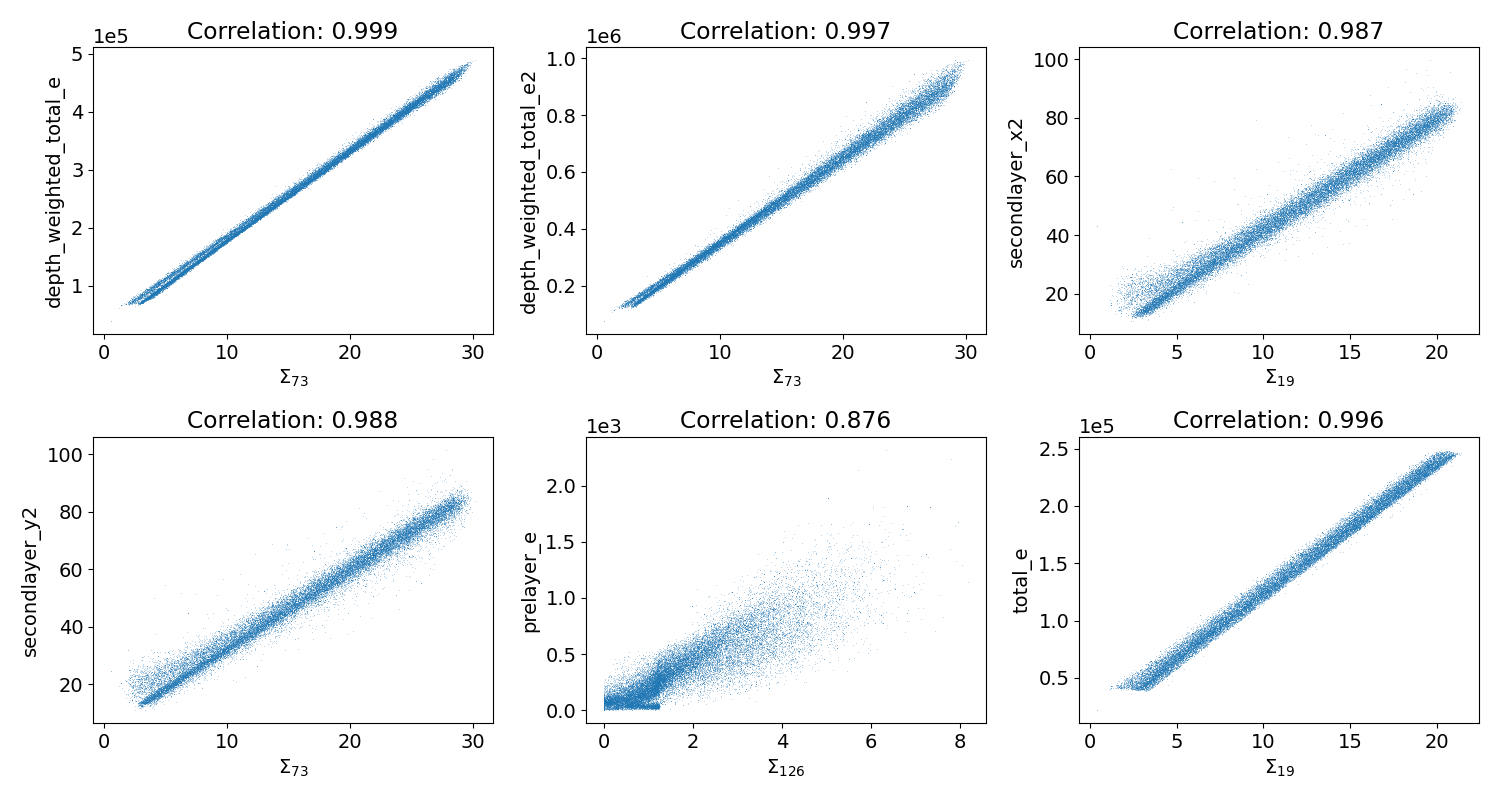}
    \caption{Visual representation of Table \ref{tab:PFN_feature_interp}. Scatter plots of the activations of six select units in the $\Sigma$ layer of our $a\rightarrow 3\pi^0$ PFN against correlated physical variables. Each point represents a single sample event.}
    \label{fig:PFN_feature_corr}
\end{figure}

Many of the units in $\Sigma$---the layer of the PFN corresponding to the sum of every particle's latent space---correlate strongly with select BDT features (input variables) which have physical interpretations. Refer to Table~\ref{tab:PFN_feature_interp} for the PFN units, their BDT feature counterparts, and the correlation value. Additionally, see Fig.~\ref{fig:PFN_feature_corr} for an analogous visual representation featuring each of the 3,000 sample events plotted against the two variables (PFN unit and physical variable).

\begin{figure}
    \centering
    \includegraphics[width=0.7\textwidth]{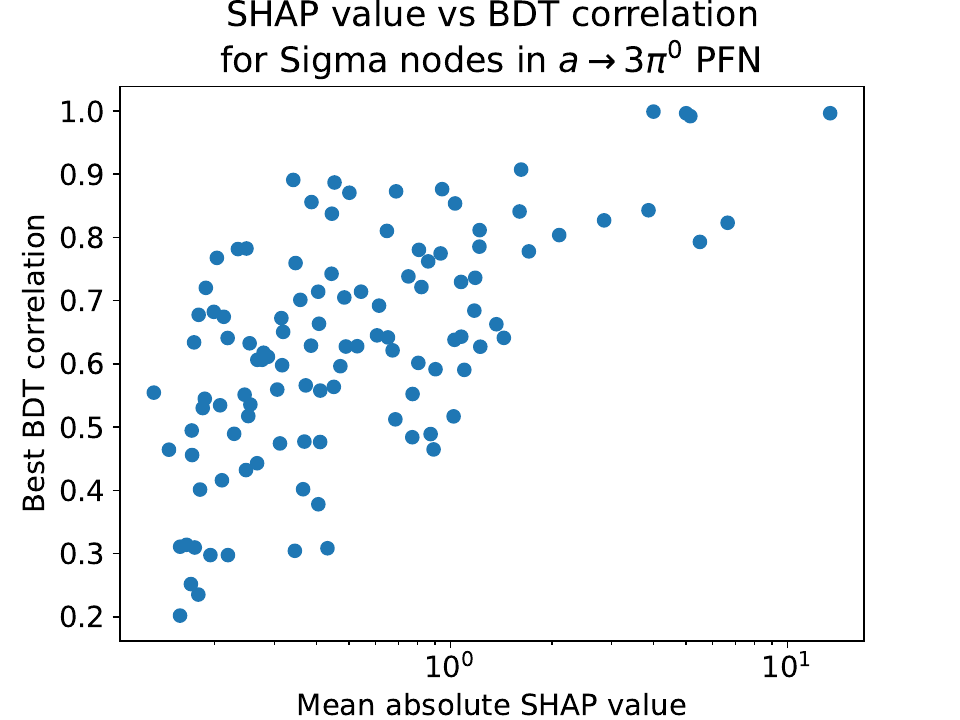}
    \caption{For each unit in the $\Sigma$ layer of our $a\rightarrow3\pi^0$ PFN, we plotted its SHAP value (how strongly it influenced the model's prediction) against its highest correlation with a BDT feature. The SHAP value axis is on a log scale.}
    \label{fig:SHAP_BDT_corr_corr}
\end{figure}

Using SHAPley values, a generalized way to explain the outputs of machine learning models using game theory, we determined that the PFN units with stronger physical variable analogs tended to have a larger impact on the model's output (see Fig.~\ref{fig:SHAP_BDT_corr_corr}). Many of the units with strongest influence on the PFN's output also correlate strongly with a BDT feature.

There are other units that have weaker, yet still moderately strong influence on the PFN's output but which don't correlate very strongly with any BDT variables. Take, for instance $\Sigma_{26}$, whose best BDT feature correlation is only $41.5\%$, but whose SHAP value of 1.04 exceeds that of other units, many with even higher BDT feature correlations. This suggests that the model learned to model other jet features not necessarily encompassed by our BDT study, and perhaps even those not directly computable from ECAL images. The PFN shows strong potential for being able to ``recover" information about the original photon-jet particle cloud that was lost in ECAL measurements.

%% Methodology: to include or not?
%% To compute these correlation values, we took $10\%$ of the photon jets this model is trained on (including axion, pion, and gamma), which amounts to ${3,\!000}$ jets. The values of the units in $F^{[7]}$ were computed for each jet, as were various physical variables. For each PFN unit, the physical variable with highest Pearson product-moment correlation was determined.

\section{Results and implication to the Higgs portal model}
\label{sec:results}

\begin{figure}
\begin{center}
\subfloat[]{\includegraphics[width=.35\columnwidth]{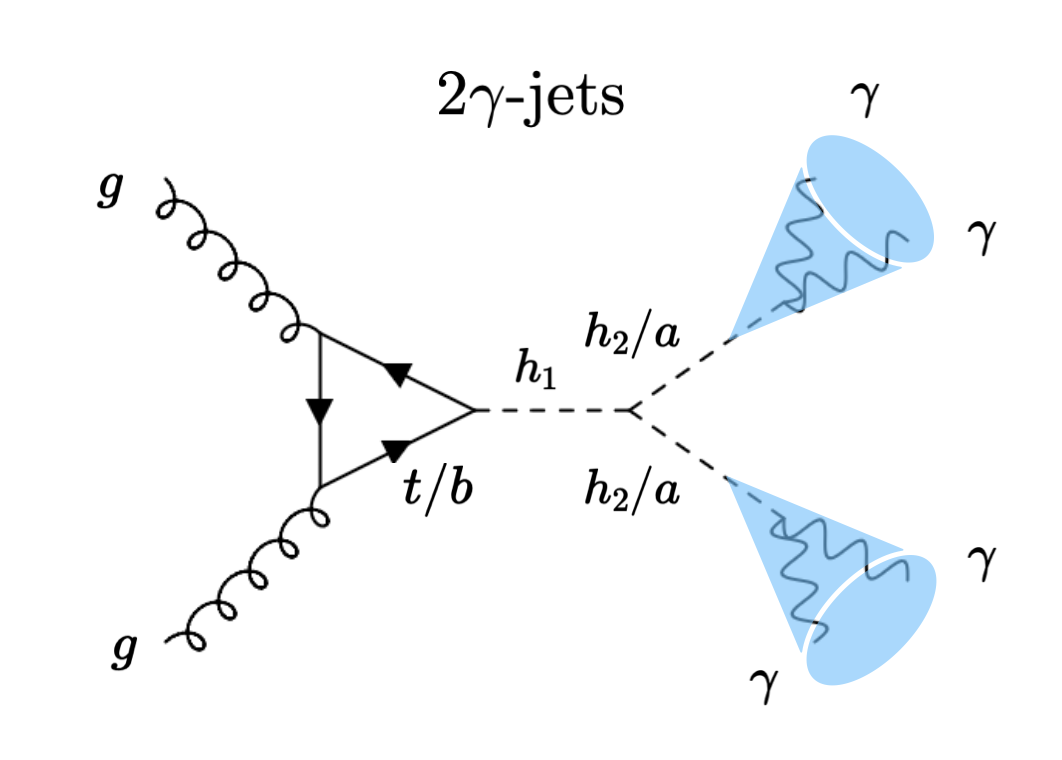}}
\subfloat[]{\includegraphics[width=.35\columnwidth]{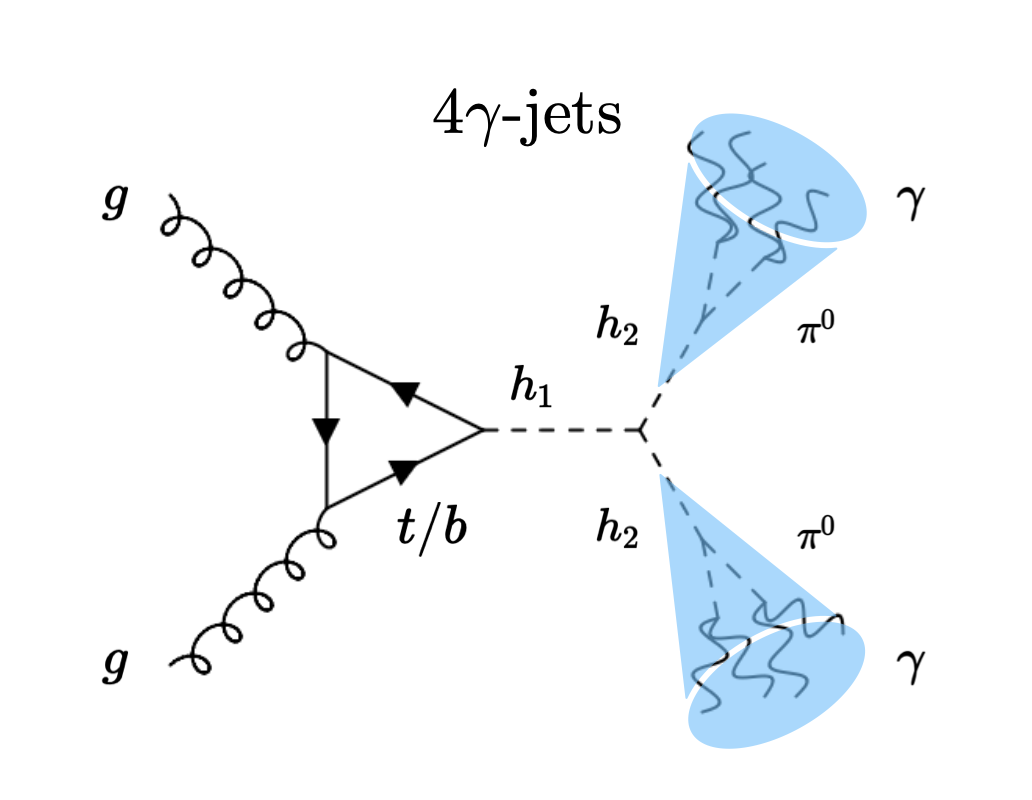}}
\subfloat[]{\includegraphics[width=.32\columnwidth]{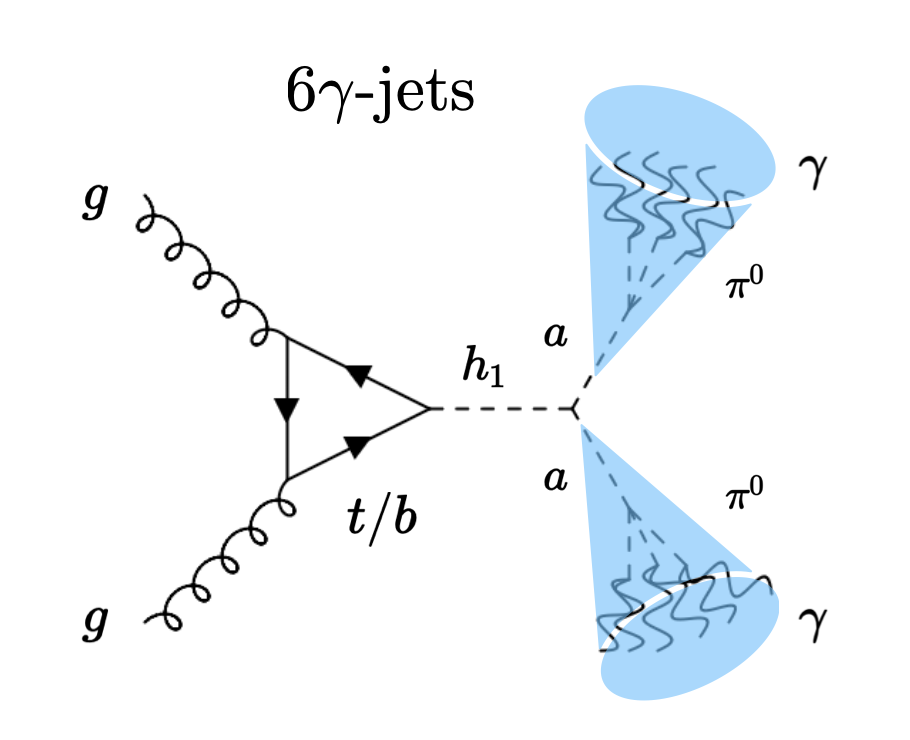}}
\end{center}
\caption{\label{fig:feyn_diag}
Feynman diagrams for photo-jet processes in the Higgs portal models. Event toplologies: (a) $gg\rightarrow h_1\rightarrow h_2 h_2/ aa\rightarrow (\gamma\gamma)(\gamma\gamma)$; (b) $gg\rightarrow h_1\rightarrow h_2 h_2\rightarrow (\pi^0\pi^0)(\pi^0\pi^0)$; (c) $gg\rightarrow h_1\rightarrow aa\rightarrow (\pi^0\pi^0\pi^0)(\pi^0\pi^0\pi^0)$. A photon-jet is denoted by each pair of parentheses.}
\end{figure}

The predominant production channel for the SM-like Higgs boson is through gluon fusion, with a cross section exceeding that of other production channels by more than an order of magnitude. In the subsequent analysis, our attention is directed specifically toward the gluon fusion production of $h_1$. Subsequently, we explore the decay process of $h_1$ into a pair of $h_2$ or $a$. Specifically, to explore the implications of the Higgs portal models in distinguishing photon-jet signatures from those of single photons and neutral pions in the SM, we investigate three distinct signal topologies: 
\begin{align} 
& gg\rightarrow h_1\rightarrow h_2 h_2/ aa\rightarrow (\gamma\gamma)(\gamma\gamma), \label{Eq:sig_1} \\ & 
gg\rightarrow h_1\rightarrow h_2 h_2\rightarrow (\pi^0\pi^0)(\pi^0\pi^0), \label{Eq:sig_2} \\ & 
gg\rightarrow h_1\rightarrow aa\rightarrow (\pi^0\pi^0\pi^0)(\pi^0\pi^0\pi^0). \label{Eq:sig_3}  
\end{align}
The corresponding Feynman diagrams for these signal processes are depicted in Fig.~\ref{fig:feyn_diag}. Notably, the highly collimated nature of the photons resulting from the decays of $h_2$ and $a$ prevents them from satisfying conventional photon isolation criteria. Consequently, novel signal signatures emerge, including $2\gamma$-jets, $4\gamma$-jets, and $6\gamma$-jets, as opposed to conventional multi-photon structures. We assume that the gluon fusion production cross section of $h_1$ is consistent with the SM prediction, denoted as $\sigma (gg\rightarrow h_1) = 54.67$ pb at $\sqrt{s}=14$ TeV~\cite{LHCHiggsCrossSectionWorkingGroup:2016ypw}. Furthermore, to preserve the model independence of our analysis, we consistently set the decay branching ratios of $h_2/a\rightarrow\gamma\gamma$, $h_2\rightarrow\pi^0\pi^0$, and $a\rightarrow 3\pi^0$ to unity for each investigation. As a result, the only two free model parameters remaining are either $\left(m_{h_2}, |\mu_{h_1 h_2 h_2}|\right)$ or $\left(m_a, |\mu_{h_1 aa}|\right)$. In this section, we intend to delineate the sensitivity of photon-jet signatures to these two model parameters individually.

We employ \textsf{FeynRules}~\cite{Christensen:2008py} to generate the UFO model files~\cite{Degrande:2011ua} for the Higgs portal models under investigation. Our analysis encompasses three major SM background processes: $gg\rightarrow h_1\rightarrow \gamma\gamma$, $pp\rightarrow\gamma\gamma$ and $pp\rightarrow\gamma j$\footnote{For the SM background $pp\rightarrow jj$, the efficiency after basic event selections is extremely low. Thus, generating a very large number of Monte Carlo background events (more than $10^{10}$) is required, which is beyond the scope of this work.}. To conduct this study, we utilize \textsf{MadGraph5\_aMC@NLO}~\cite{Alwall:2014hca} with the NN23LO1 PDF set~\cite{NNPDF:2017mvq} to simulate the leading-order contributions of the signal processes described in Eqs.~(\ref{Eq:sig_1}) to (\ref{Eq:sig_3}) and the three SM background processes occurring in $pp$ collisions at $\sqrt{s}=14$ TeV. The cross section for the gluon fusion production  of $h_1$ has been rescaled with N$^3$LO QCD and NLO EW accuracies~\cite{LHCHiggsCrossSectionWorkingGroup:2016ypw}. We applied ME-PS matching with the MLM prescription~\cite{Mangano:2006rw,Alwall:2007fs} to $pp\rightarrow\gamma\gamma$ and $pp\rightarrow\gamma j$, including the emission of up to two additional partons for these two processes. Additionally, Monte Carlo events are generated from MadGraph5\_aMC@NLO for both the signal and background processes. All events undergo parton showering, hadronization, and the treatment of unstable particle decays through \textsf{Pythia8}~\cite{Bierlich:2022pfr}. Our simulation culminates with the identification of photon-jet signatures originating from either $h_2$ or $a$, a task carried out using an ATLAS-like ECAL. Notably, the electromagnetic showers are simulated with GEANT4, as outlined in Sec.~\ref{sec:identification}. 
%Moreover, this study accounts for the pileup effect, a crucial consideration for a comprehensive analysis at the LHC. 

Following the Pythia8 simulations, we apply specific criteria to the generated truth-level events for both signal and background processes in order to identify photon-jet candidates. These criteria are intended to select events featuring at least two photon-jet candidates, which are defined as follows. We employ the anti-$k_T$ jet clustering algorithm~\cite{Cacciari:2008gp} with $R_J < 0.25$ to group collimated photons, where $R_J$ denotes the jet cone radius. Given that photon-jet candidates primarily deposit their energy in the ECAL rather than the HCAL associated with ordinary QCD jets, we require $\log\theta_J < -0.8$, where $\theta_J$ represents the hadronic energy fraction of a jet. Within the truth-level events, the hadronic energy fraction encompasses all charged and neutral hadrons, excluding $\pi^0$ and $\eta$ mesons in our analysis. Both $\pi^0$ and $\eta$ will dominantly decay to a pair of photons, which deposit almost all of their energy in the ECAL instead of the HCAL. Furthermore, within the jet cone of $R_J < 0.25$, we veto all charged tracks with $P_T > 2$ GeV to mitigate the possibility of QCD jets mimicking photon-jet candidates. On the other hand, to enforce a loose track isolation criterion for the photon-jet candidates, they must be isolated from nearby charged tracks within a cone radius of $\Delta R < 0.2$, where the scalar sum of the $P_T$ of tracks is required to be smaller than $5\%$ of the photon-jet candidate's $P_T$. Since two photon-jet candidates originate from the SM-like Higgs boson through gluon fusion in this study, they are expected to be energetic and distributed in the central region. Consequently, we require that the leading photon-jet, $J_1$, and the sub-leading photon-jet, $J_2$, meet the criteria of $P_T(J_{1,2}) > 40$ GeV and $|\eta_J| < 2.5$. Additionally, we utilize the invariant mass distribution of the two photon-jet candidates ($M_{J_1 J_2}$), which can be reconstructed to align with the SM-like Higgs boson mass in the signal processes and $gg\rightarrow h_1\rightarrow \gamma\gamma$. However, this distribution displays a smooth, decreasing profile for $pp\rightarrow\gamma\gamma$ and $pp\rightarrow\gamma j$. To bolster the suppression of SM background events while preserving most signal events, we apply the following event selection criteria: $P_T (J_1) > 0.4 M_{J_1 J_2}$, $P_T (J_2) > 0.3 M_{J_1 J_2}$, and $120$ GeV $< M_{J_1 J_2} < 130$ GeV.

Following the event selections of the photon-jet candidates as described above at the truth level, we employ advanced machine learning techniques to further differentiate these photon-jet signatures from those originating from SM background events. As elucidated in Sec.~\ref{sec:identification}, both CNN and PFN emerge as the potent tools for the discrimination of photon-jet signatures and the rejection of SM background events. We consequently utilize the identification efficiencies for photon-jet signatures and SM backgrounds achieved through the CNN as an example to determine the event counts at the reconstructed level. It is important to note that these efficiencies are assumed to be independent of the variable $\eta$. Given that our two photon-jet candidates are predominantly distributed in the central region, this assumption remains valid for our analysis. 
%{\color{red}(Discuss the pileup effect here !)} 

\begin{figure}
\begin{center}
\subfloat[]{\includegraphics[width=.5\columnwidth]{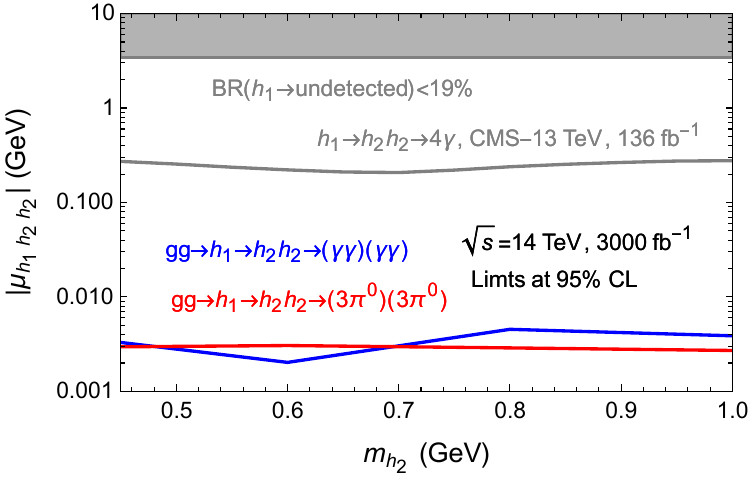}}
\subfloat[]{\includegraphics[width=.5\columnwidth]{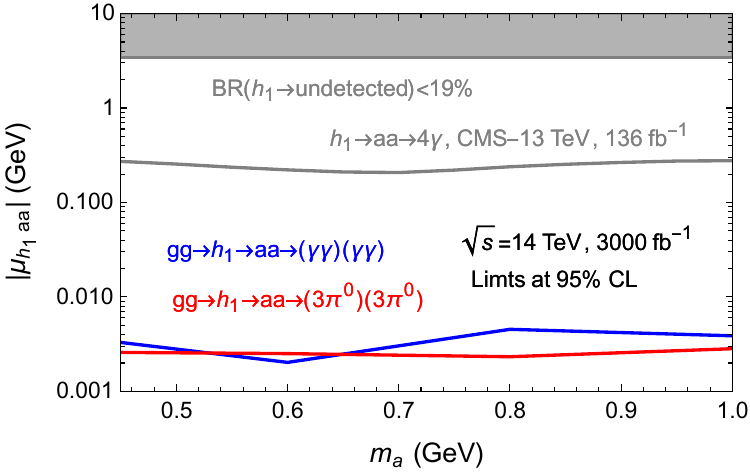}}
\end{center}
\caption{\label{fig:sensitivity} The projection of limits at $95\%$ CL for (a) $(m_{h_2}, |\mu_{h_1 h_2 h_2}|)$ and (b) $(m_a, |\mu_{h_1 aa}|)$ for various photon-jet signatures in Higgs portal models at the LHC with $\sqrt{s}=14$ TeV and an integrated luminosity of $3000\,\textrm{fb}^{-1}$ without the pileup effect. The top gray bulk is the constraint from the exotic decay branching ratios of the SM-like Higgs boson, $B(h_1\rightarrow\text{undetected}) < 19\%$~\cite{ATLAS:2020qdt} and the gray line is the constraint from the CMS search ($13$ TeV, $136$ fb$^{-1}$) of exotic Higgs boson decays $h_1\rightarrow h_2 h_2/aa\rightarrow 4\gamma$~\cite{CMS:2022wpu}, respectively.}
\end{figure}

We can then translate these acceptances, derived from the event selections and CNN identification efficiencies, into constraints on the Higgs portal model parameters $\left(m_{h_2}, |\mu_{h_1 h_2 h_2}|\right)$ and $\left(m_a, |\mu_{h_1 aa}|\right)$, depending on whether the light scalar is CP-even or CP-odd. The projections of the $95\%$ confidence level (CL) limits for $|\mu_{h_1 h_2 h_2}|$ (left) and $|\mu_{h_1 aa}|$ (right) within the range $0.45$ GeV $\leqslant m_{h_2, a}\leqslant 1.0$ GeV, without the pileup effect, for $\sqrt{s} = 14$ TeV and an integrated luminosity of ${\cal{L}} = 3000$ fb$^{-1}$, are presented in Fig.~\ref{fig:sensitivity}. For the sake of comparison, we incorporate the constraint from the exotic decay branching ratios of the SM-like Higgs boson, $B(h_1\rightarrow\text{undetected}) < 19\%$~\cite{ATLAS:2020qdt}, which arises from data at $\sqrt{s} = 13$ TeV and an integrated luminosity of ${\cal{L}} = 139$ fb$^{-1}$, into Fig.~\ref{fig:sensitivity}. Here $B(h_1\rightarrow\text{undetected})$ is defined as $\Gamma_{\text{BSM}}/\left(\Gamma_{\text{BSM}}+\Gamma_{\text{SM}}\right) $ with $\Gamma_{\text{BSM}} = \Gamma (h_1\rightarrow h_2 h_2)$ or $\Gamma (h_1\rightarrow aa)$ and $\Gamma_{\text{SM}}=4.03$ MeV~\cite{LHCHiggsCrossSectionWorkingGroup:2011wcg}. This constraint can be translated to $|\mu_{h_1 h_2 h_2}|\lesssim 3.34$ GeV and $|\mu_{h_1 aa}|\lesssim 3.34$ GeV, respectively. 
%{\color{red}(Some comments about Fig.1 and Fig.2 should be added here!)} 

Finally, we conduct a comparison of our findings with previous photon-jet searches conducted by both the ATLAS and CMS Collaborations. The ATLAS Collaboration has searched for similar signatures in three distinct publications~\cite{ATLAS:2012soa,ATLAS:2018dfo,ATLAS:2023eet}. In the first one~\cite{ATLAS:2012soa}, the investigation centers around Higgs boson exotic decays to pairs of light pseudoscalars with masses below $0.4$ GeV. This scenario involves largely boosted pseudoscalars, making the final state challenging to distinguish from a single-photon signature, thus serving as a valuable complement to our study. The second one~\cite{ATLAS:2018dfo} explores high-mass resonance scalar particles with masses exceeding $200$ GeV, differing from our focus on the SM-like Higgs boson mass. Consequently, a direct comparison with our results is not feasible. In the third one~\cite{ATLAS:2023eet}, the investigation involves Higgs boson exotic decays to pairs of axion-like particles with masses exceeding $5$ GeV, providing another complementary aspect to our study. The CMS Collaboration, in their study~\cite{CMS:2022wpu}, explores Higgs boson exotic decays to pairs of light scalars with masses ranging from $0.1$ GeV to $1.2$ GeV, aligning with the scope of our research. Consequently, we incorporate these constraints into Fig.~\ref{fig:sensitivity} for comparison. Notably, our advanced machine learning techniques enable the exploration of much smaller values (${\cal{O}}(1)$ MeV) for $|\mu_{h_1 h_2 h_2}|$ and $|\mu_{h_1 aa}|$ at the High-Luminosity LHC. 

\section{Conclusion}
\label{sec:conclusion}
%% Conclusion 
The Higgs boson, the final component of the Standard Model (SM), was discovered over ten years ago. Various properties of this SM-like Higgs boson have been precisely measured, including its mass, width, charge-parity (CP), and couplings to gauge bosons and the third generation of fermions. In future studies, measurements of the Higgs boson self-coupling, its coupling to the second generation of fermions, and other properties can shed light on the origin of electroweak symmetry breaking and the mass origin of matter. Beyond examining these fundamental properties of the scalar boson, a crucial aspect involves directly investigating its potential connection to new physics. This linkage between the SM sector and the dark sector is known as the Higgs portal. In models where new particles have masses less than half of the SM-like Higgs boson within the Higgs portal, the Large Hadron Collider (LHC) provides an opportunity to explore exotic and invisible decays of the Higgs boson. Therefore, the search for these potential new decay channels of the SM-like Higgs boson at the LHC represents a powerful approach to unravel the mysteries of new physics.

We specifically investigate a particular aspect of Higgs boson exotic decays, focusing on the photon-jet signatures at the LHC in this study. The unique photon-jet signautre arises when a few collimated photons are generated, failing to satisfy the standard photon isolation criteria. If a pair of new light particles is produced from the SM-like Higgs boson, primarily decaying to photons in the final state, two highly boosted light particles result in the natural generation of two photon-jets. Due to the photon-jet's predominant energy deposition in the electromagnetic calorimeter (ECAL), it exhibits behavior akin to a single photon and neutral pion in the SM. To discern the photon-jet signature from SM backgrounds, we employ advanced machine learning techniques in this investigation. This work encompasses three photon-jet signals: $h_2/a\rightarrow\gamma\gamma$, $h_2\rightarrow\pi^0\pi^0$, and $a\rightarrow 3\pi^0$, where $h_2$ ($a$) represents a light scalar (pseudoscalar). Our findings, depicted in Fig.~\ref{fig:PFN_CNN_comp_eff}, indicate that both Convolutional Neural Networks (CNN) and Particle Flow Networks (PFN) serve as potent tools for distinguishing photon-jet signatures from SM backgrounds, such as the single photon and neutral pion from QCD jets. Taking CNN as an example, the photon-jet can be identified with an efficiency exceeding $99\%$, accompanied by a background rejection rate surpassing $99\%$. Notably, our comparative analysis reveals that the two deep learning methodologies, CNN and PFN, outperform the traditional machine learning analysis, Boosted Decision Trees (BDT), as illustrated in Fig.~\ref{fig:roc_curves} and Table~\ref{tab:eff_compare}.

With the aid of basic event selections and deep learning analysis, we can forecast future constraints on the $h_1-h_2-h_2$ and $h_1-a-a$ three-point interactions within Higgs portal models, as denoted by the couplings $\mu_{h_1 h_2 h_2}$ and $\mu_{h_1 aa}$, both possessing dimensions equivalent to mass. Notably, investigating exotic decays of the Higgs boson proves to be the most suitable approach for detecting the magnitudes of these three-point interactions. Projections of the $95\%$ Confidence Level (CL) limits for $|\mu_{h_1 h_2 h_2}|$ and $|\mu_{h_1 aa}|$ at the High-Luminosity LHC, without pileup effects, are presented in Fig.~\ref{fig:sensitivity}. Our findings reveal that $2\sigma$ bounds for $|\mu_{h_1 h_2 h_2}|$ and $|\mu_{h_1 aa}|$ can be explored down to the order of ${\cal{O}}(1)$ MeV, demonstrating the efficacy of CNN and PFN in achieving this level of precision.

%% Acknowledgements
\acknowledgments
The work of S.-C. Hsu and K. Li are supported by the U.S. Department of Energy, Office of Science, Office of Early Career Research Program under Award number DE-SC0015971. The work of C.-T. Lu is supported by the National Natural Science Foundation of China (NNSFC) under grant No. 12335005.

% Bibliography
%% Using JHEP.bst file
%%%\bibliographystyle{JHEP}
%%%\bibliography{references.bib}

\end{document}